# Hurricane Effects on Mangrove Canopies Observed from MODIS and SPOT Imagery

## Michael S. Parenti

January 2015

Strategy & Analytics, Saatchi & Saatchi Wellness

New York, NY 10014

thisismikep@gmail.com

**Abstract.** The effects of two hurricanes (Katrina and Wilma) on protected mangroves in southwest Florida and two hurricanes (Emily and Dean) on protected mangroves in the Yucatan Peninsula were assessed paired sets of 20 m multispectral SPOT and 16-day 500m MODIS images. Normalized difference vegetation index (NDVI) and enhanced vegetation index (EVI) values were calculated to measure mangrove canopy reflectance using three methods of analysis— 1) mangrove NDVI sample point extraction, 2) qualitative assessment of 500m MODIS NDVI and EVI time series, and 3) transects of NDVI differences before and after storm events. Results show each method effectively captures changes in mangrove canopies consistent with storm effects. However, the extent of damage to mangroves in South Florida and Sian Ka'an was highly variable. Hurricanes Wilma and Katrina produced a large drop in NDVI although NDVI values recovered within a year, suggesting remarkable resilience. Hurricane Emily apparently increased mangrove photosynthetic activity owing to the location of landfall relative to the study area, the size of the wind field and the apparent lack of storm surge. Comparison of SPOT NDVI and MODIS NDVI values revealed that the changes pre- and post-storms were consistent at the different scales of observation.

## 1 INTRODUCTION

Mangrove and adjacent wetland ecosystems are among the most productive ecosystems on earth [1] but are also among the most vulnerable to anthropogenic activities such as dredge and fill, aquaculture, construction, and other environmental modifications in tropical coastal zones [2]. Mangrove ecosystems also provide many important functions that enhance overall productivity of estuarine and coastal fisheries; they stabilize and may protect coastlines from erosion, and reduce storm and wave impacts in coastal regions. In addition to direct threats posed by coastal development, future atmospheric warming trends coupled with sea-level rise are likely to exacerbate existing threats caused by local anthropogenic disturbance [2].

Biogeographic studies reveal that mangroves canopy height is inversely related to latitude, with taller trees possessing higher biomass in equatorial locations relative to stands found closer to the subtropics [3]. Among the plausible reasons advanced for this geographic trend are land-falling tropical cyclones (known as hurricanes in the Atlantic Basin), which cause widespread mortality of mangrove trees [1, 4-5]. In the western hemisphere, hurricanes tend to occur most frequently in the northern Caribbean, Gulf of Mexico, and South Florida [6]. In 2004 and 2005 this region experienced an anomalously high number of tropical storms and hurricanes, which may be explained by positive warming trends in sea surface temperature [7].

In the Yucatan's Sian Ka'an Biosphere Reserve and South Florida's Everglades National Park, several factors influence the persistence and productivity of mangrove communities where they are protected from coastal development. A number of studies have examined the influence of tropical storms and hurricanes on mangrove community resilience and resistance to these major disturbance events [4-5, 8]. Owing to their position along coastal margins, hurricane-related mortality in mangrove forests may be high (> 30 percent); however, mangrove trees seem to possess significant resilience if storm frequencies are sufficiently low to allow regeneration through re-sprouting and seedling growth [4-8].

Application of optical remote sensing to study mangrove environments has grown over the past decade [9-13]. Remote sensing studies have revealed much about mangrove ecosystems including canopy heights, leaf densities, deforestation, erosion, and pollution and other anthropogenic changes. A number of remote sensing studies have employed multispectral satellite data from Landsat and SPOT, which provide resolutions appropriate for studying changes in mangrove canopy extent and condition. More recently, higher spatial and spectral resolution systems have been used for analysis of within-stand variability and changes [14-15]. However, the cost of using new high-resolution systems remains high and these systems tend to provide limited spatial and temporal coverage relative to medium-to-low spatial resolution systems.

The Normalized Difference Vegetation Index (NDVI) and mangrove biomass hold a strong linear relationship [16].  Muttitanon and Tritracki [17] showed that NDVI image differencing, a method of subtracting one date of NDVI imagery from another, provides an indirect measure of mangrove leaf biomass and is a useful tool for mapping the areal extent.  The NDVI formula is calculated as:

$$NDVI = \frac{NIR - RED}{NIR + RED} \tag{1}$$

where NIR (Near Infrared) and RED (Red) are spectral reflectance measurements acquired in the red and near-infrared regions. Negative values of NDVI (values approaching -1) correspond to water, snow, ice and some dark surfaces such as burn scars. Positive values (approaching +1) are linearly related to photosynthetically active trees, shrub, grassland, and other forms of vegetation [18-19].

Two major limitations of NDVI include 1) soil backscatter from low-lying vegetation and 2) saturation from dense vegetation.  The Enhanced Vegetation Index (EVI) was designed to reduce these effects, since it is resistant to aerosol scattering and soil background, and does not saturate when viewing dense vegetation and other areas with large amounts of chlorophyll [18].  EVI is given by [20]:

$$EVI = G\left(\frac{(NIR/RED) - 1}{(NIR/RED) + \left[C_1 - C_2 \times (BLUE/RED)\right] + (L/RED)}\right) \tag{2}$$

where *L* is a soil adjustment factor, and $C_1$ and $C_2$ are coefficients used to correct aerosol scattering in the red band by the use of the blue band. The *blue*, *red*, and *NIR* represent reflectance at the blue (0.45-0.52μm), red (0.6-0.7μm), and NIR wavelengths (0.7-1.1μm), respectively.

The purpose of this study is to a) evaluate the impact of major storms on the Sian Ka'an and Everglades mangrove ecosystems, which have a history of storm impacts, b) to examine storm resilience and c) to investigate the phenomena at different spatial and temporal scales using optical sensors from two different orbital platforms, SPOT and MODIS.

## 2 STUDY AREAS

### 2.1 Sian Ka'an Biosphere Reserve, Quintana Roo, Mexico

Located on the Caribbean Coast in the State of Quintana Roo, the 6,510 km$^2$ Sian Ka'an Biosphere Reserve is one of Mexico's largest protected areas. Offshore barrier coral reefs are separated from large interior freshwater wetlands by extensive, highly productive mangrove estuaries and lagoons, nearly 90,000 ha [21]. The area is almost entirely monospecific with red mangrove, *Rhizophora mangle*. Black mangrove, *Avicennia germinans*, tonga mangrove, *Lumnitzera racemosa* and button wood, *Conocarpus erectus* are very rare [22]. The asymmetrical ebb and flood tides, with the ebb tide being shorter but with stronger current velocity than the flood tide, is altered during tropical storms and hurricanes allowing for defoliation, uprooting, and erosion of shallow roots to occur [23].  As a natural line of defense, the Sian Ka'an mangrove ecosystem is largely protected from the sea by a 110 kilometre-long 15,000 ha stretch of the 1,200 km Mesoamerican reef.  This is the second largest barrier reef in the world, the growth of which is partly due to the lack of erosion inland and consequent silt-free water [24].  During the last century there has been an average of one hurricane every eight years.  The most recent land-falling tropical cyclone was Dean in 2007.  Fig. 1 shows the location of the study area with the tracks of Dean and Emily superimposed.

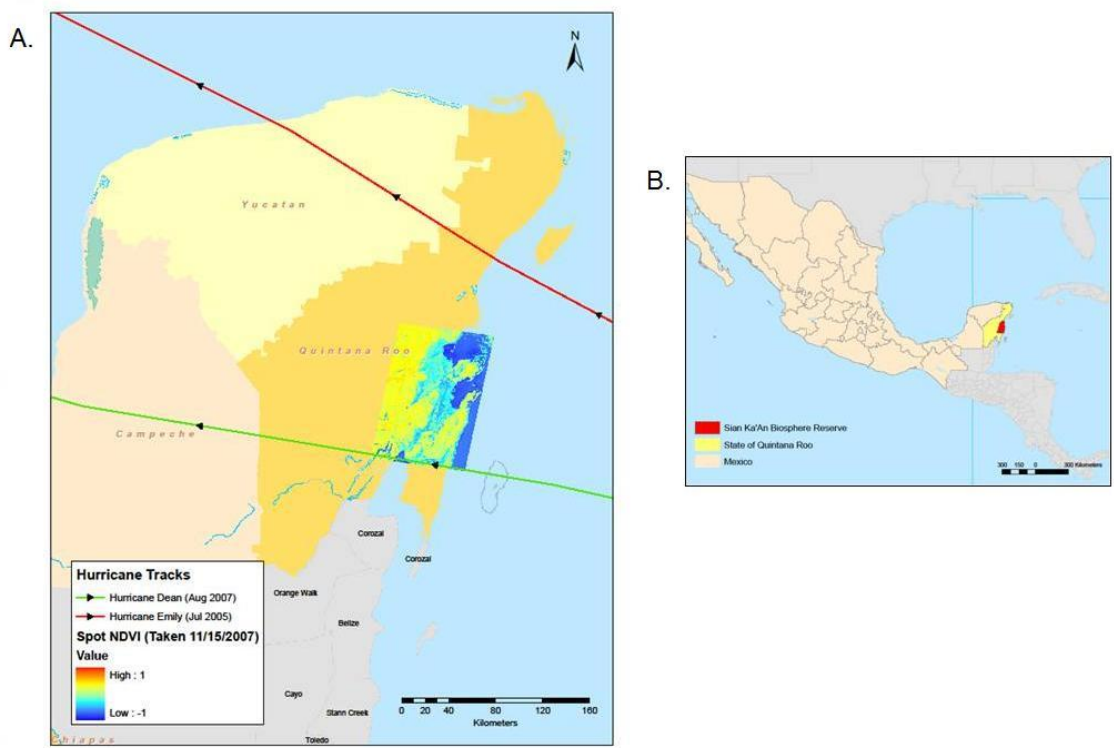

Fig. 1.  A. Location of the study area in Quintana Roo, Mexico showing an overlay of SPOT false color imagery and the two hurricane tracks (Hurricanes Dean and Emily) over the study area. Hurricane tracks from http://www.nhc.noaa.gov.  B. Study area with regards to the region.

## 2.2 Everglades National Park, Florida, USA

The Everglades National Park is home to the largest mangrove ecosystem in the western hemisphere [25].  It contains three species of mangrove: red mangrove (*Rhizophora mangle*), black mangrove (*Avicennia germinans*) and white mangrove (*Laguncularia racemosa*).  In addition, *Conocorpus erecta* is commonly found as an associate of these three species.  These species are found in mixed stands or in zones on the subtropical southwestern coast of Florida, which has a very gentle westward slope into the Gulf of Mexico (Fig. 2).

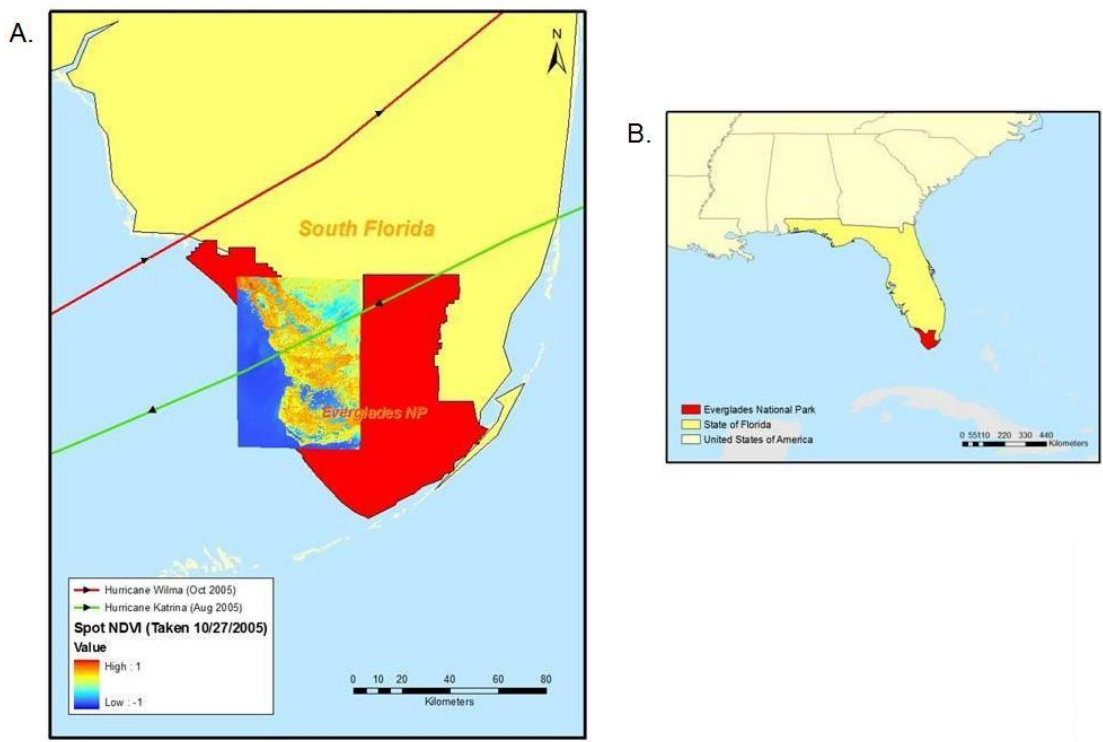

Fig. 3.  A. Location of the study area in southwest Florida, USA showing an overlay of SPOT false color imagery and the two hurricane tracks (Hurricanes Katrina and Wilma)

over the study area. Hurricane tracks from http://www.nhc.noaa.gov.  B. Study area with regards to the region.

In the Everglades, hurricanes occur between June and November and are a persistent threat to woody shoreline vegetation.  Mangrove species are susceptible to disturbance by hurricanes such as the 2005 hurricanes (Katrina and Wilma) that caused significant changes in forest structure and function [26].  Fig. 3 shows the location of the Everglades National Park with the track of Katrina and Wilma superimposed.  Field observations made four months after the passage of Wilma revealed that the hurricanes produced partial-to-complete defoliation and much damage to woody canopy components (Fig. 3).  Some of the observed damage may have been due to the storm surge from Hurricane Wilma, which exceeded two meters along parts of the coastal zone [27].

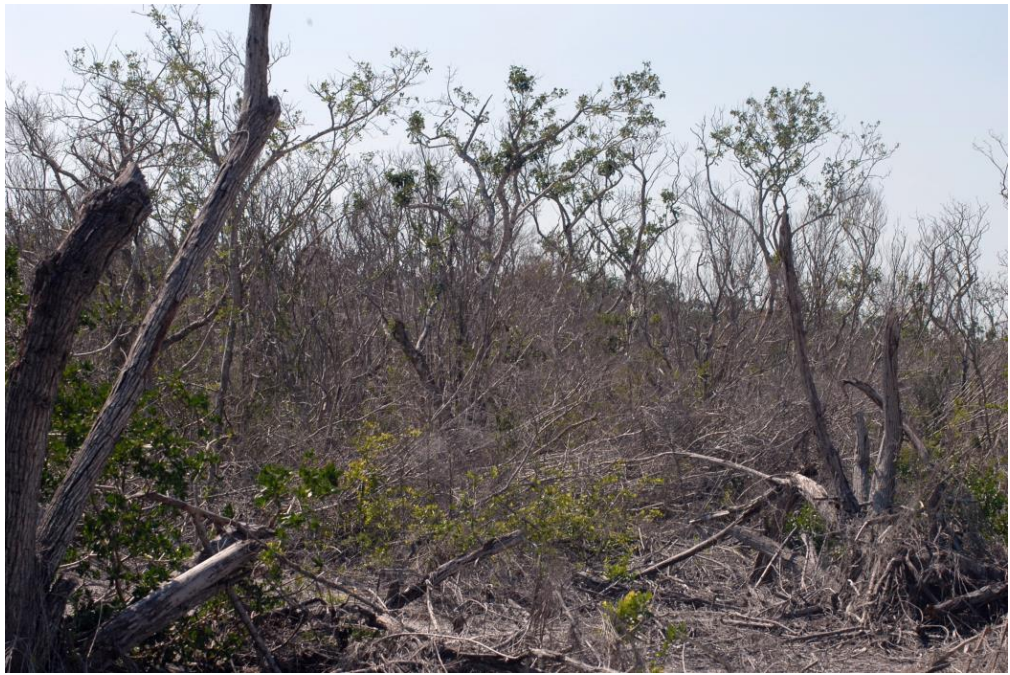

Fig. 3. Photograph of hurricane damage to mangrove canopies taken in March 2006, approximately five months after Wilma.  Photo credit: D.O. Fuller.

## 3 HURRICANE EVENTS ANALYZED

Four hurricanes, Dean and Emily for the Sian Ka'an and Katrina and Wilma for the Everglades, were selected to analyze in the study.

### 3.1 Hurricane Dean (13-23 August 2007)

Dean was a classic Cape Verde cyclone that moved through the Caribbean as a major hurricane, passing very close to Jamaica and later making landfall on the east coast of the Yucatan Peninsula as a category 5 hurricane on the Saffir-Simpson scale, which classifies hurricanes based on wind speed, barometric pressure and potential damage to vegetation and built structures.  At the time of landfall, Dean is estimated to have had a minimum central pressure of 905 mb and maximum sustained winds of 150 kt [28].  Fig. 4 shows Dean moving westward after making landfall near the center of the Yucatan Peninsula, with the radius of maximum wind covering the Sian Ka'an Biosphere Reserve

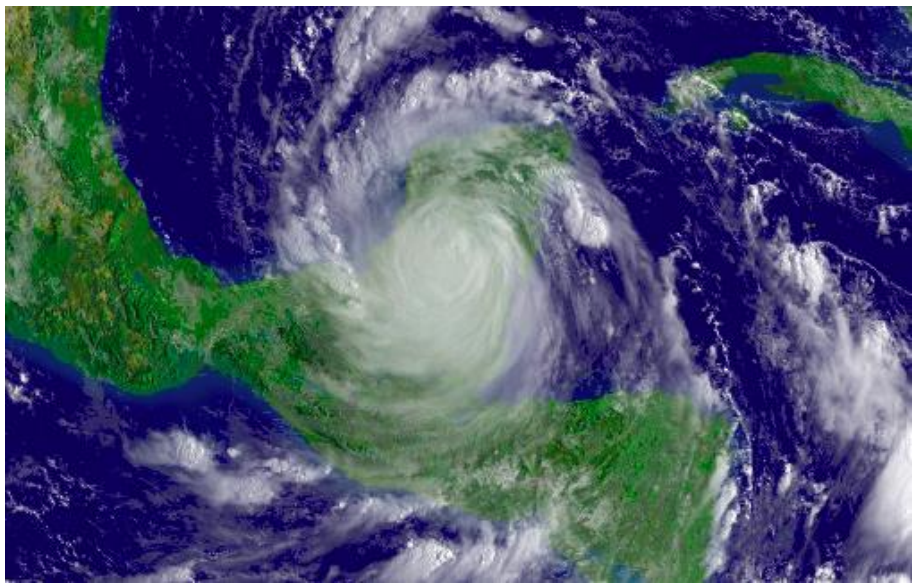

Fig. 4. Hurricane Dean regional imagery, 2007.08.21 at 1415Z. Centerpoint Latitude: 10:04:06N Longitude: 89:24:05W.   Optical imagery from NOAA Environmental Visualization Laboratory (see http://www.nnvl.noaa.gov/).

## 3.2 Hurricane Emily (11-21 July 2005)

Emily was briefly a category 5 hurricane (on the Saffir-Simpson Hurricane Scale) in the Caribbean Sea that, at lesser intensities, struck Grenada, resort communities on Cozumel and the Yucatan Peninsula, and northeastern Mexico just south of the Texas border. Emily is the earliest-forming category 5 hurricane on record in the Atlantic basin and the only known hurricane of that strength to occur during the month of July.  Emily gained winds of 115 kt, when the eye wall passed over Cozumel and when the center made landfall on the Yucatan peninsula near Tulum at 0630 UTC 18 July.  Official rainfall totals on the Yucatan were generally close to 1 inch [29].  Fig. 5 shows Emily moving westward just before making landfall near the north most area of the Yucatan Peninsula, with the radius of maximum wind just intersecting the northern most part of the Sian

Ka'an Biosphere Reserve.

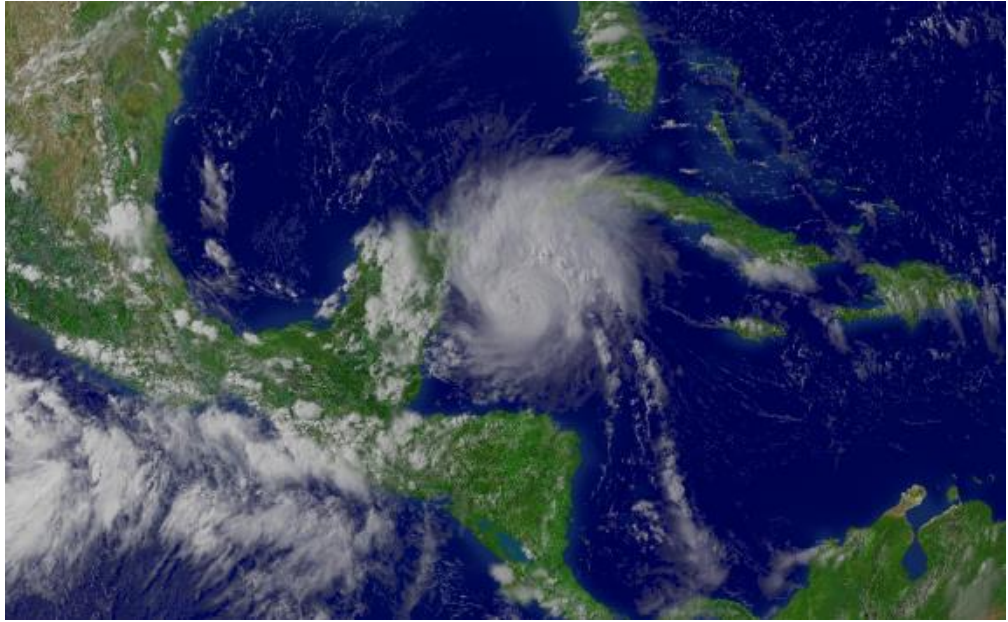

Fig. 5. Hurricane Emily regional imagery, 2005.07.17 at 2115Z. Centerpoint Latitude: 19:46:14N Longitude: 85:17:55W. Optical imagery from NOAA Environmental Visualization Laboratory (see http://www.nnvl.noaa.gov/).

## 3.3 Hurricane Katrina (23-30 August 2005)

Katrina was an extraordinarily powerful and deadly hurricane that carved a wide swath of catastrophic damage and inflicted large loss of life. It was the costliest and one of the five deadliest hurricanes to ever strike the United States. Katrina first caused fatalities and damage in southern Florida as a Category 1 hurricane on the Saffir-Simpson Hurricane Scale. Katrina made landfall near the border of Miami-Dade County and Broward County. It continued west-southwestward overnight and spent about six hours over land, mostly over the water-laden Everglades. Surface observations and velocity estimates from the Miami and Key West Doppler radars indicated that Katrina weakened over mainland Monroe County to a tropical storm with maximum sustained winds of 60 knots [30]. Fig. 6 shows Katrina moving westward crossing South Florida, with the radius of maximum wind covering the Everglades National Park.

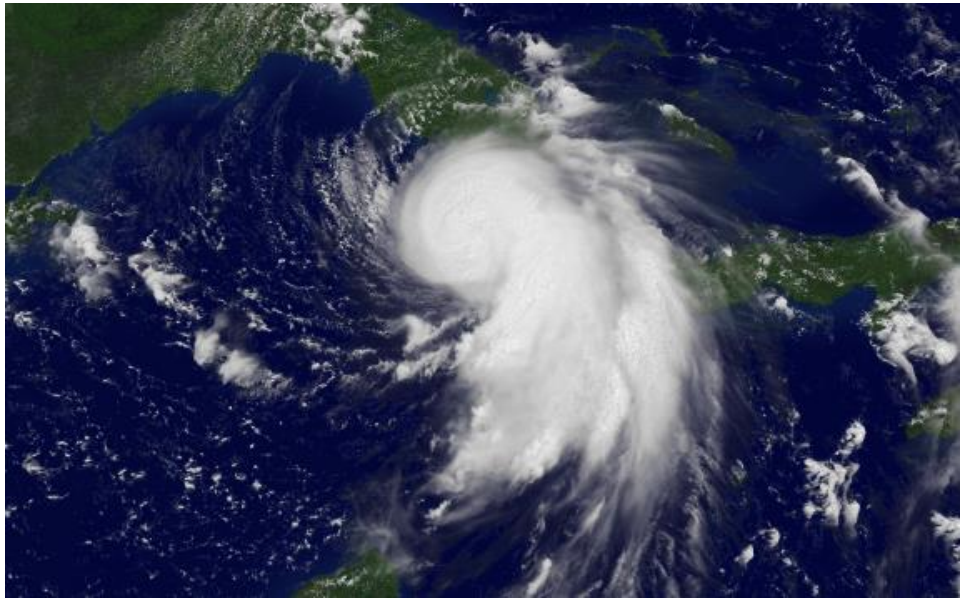

Fig. 6. Hurricane Katrina regional imagery, 2005.08.26 at 1715Z. Centerpoint Latitude: 24:34:19N Longitude: 83:17:44W. Optical imagery from NOAA Environmental Visualization Laboratory (see http://www.nnvl.noaa.gov/).

### 3.4 Hurricane Wilma (15-25 October 2005)

Wilma formed and became an extremely intense hurricane over the northwestern Caribbean Sea. It had the all-time lowest central pressure for an Atlantic basin hurricane, and it devastated the northeastern Yucatan Peninsula before inflicting extensive damage over the Everglades National Park. Based on the surface observations and the Doppler data it can be concluded that most of the southeastern Florida peninsula experienced at least category 1 hurricane conditions [31].  Fig. 7 shows Wilma moving eastward crossing South Florida, with the radius of maximum wind covering the Everglades National Park.  Tropical Depression Alpha also prepares to be absorbed by Hurricane Wilma.

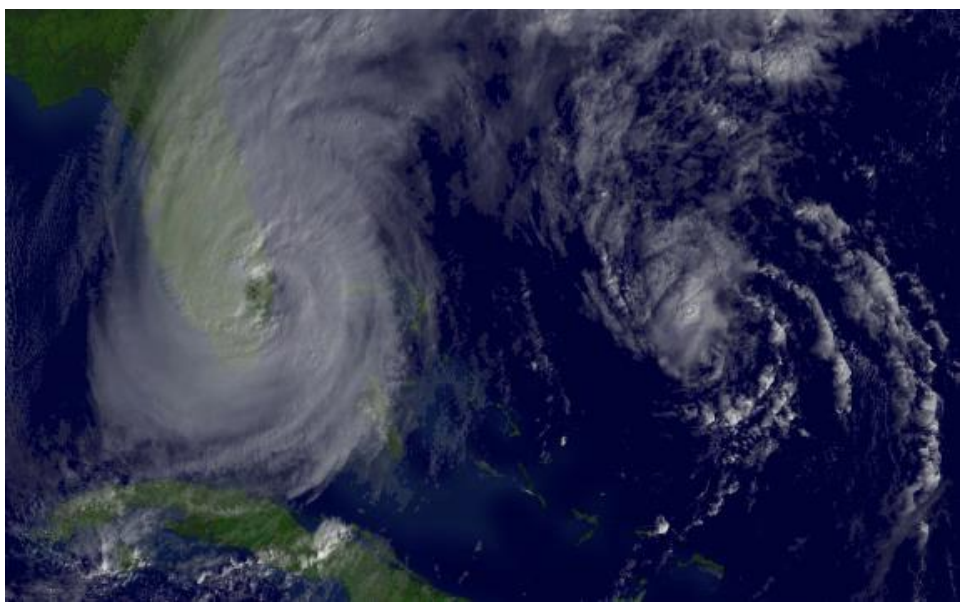

Fig. 7. Hurricane Wilma, Tropical Depression Alpha regional imagery, 2005.10.24 at 1345Z. Centerpoint Latitude: 26:27:24N Longitude: 76:42:39W. Optical imagery from NOAA Environmental Visualization Laboratory (see http://www.nnvl.noaa.gov/).

# 4 METHODOLOGY

## 4.1 Hurricane selection

The most important criterion for hurricane selection was proximity of the hurricane track to the study area.  A boundary polygon of the Sian Ka'an Biosphere Reserve was digitized based on a georeferenced map provided by Amigos de Sian Ka'an, a non-profit conservationist organization funded through the National Council of Science and Technology and the World Wildlife Fund (WWF).  A centroid of the boundary polygon was calculated by adding two new fields (x and y) to the polygon's attribute table.

Hurricane destructive winds and rains cover a wide swath. Hurricane-force winds can extend outward to about 50 km from the storm center of a small hurricane and exceed 150 km for a large one [32].  With the centroid of the Sian Ka'an study site calculated a median 100 km buffer was created around the centroid to define the area that a hurricane track must intersect to be considered for selection.  Hurricane track data, provided by the National Oceanic and Atmospheric Administration (NOAA) Coastal Service Center, was overlaid on the 100 km buffer.

The most recent hurricanes to fall within the Sian Ka'an buffer included Hurricanes Roxanne (1995, category 3), Dolly (1996, category 1), Emily (2005, category 4) and Dean (2007, category 5).  High-resolution Landsat imagery was acquired for each of the hurricanes through the University of Maryland's Global Landcover Facility free of charge.  Data integrity issues arose from a failure of the scane line corrector (SLC), so that all imagery collected after June 2003 were SLC-off resulting in data gaps (Fig. 8).

Landsat imagery was discarded and replaced with high-resolution SPOT imagery.

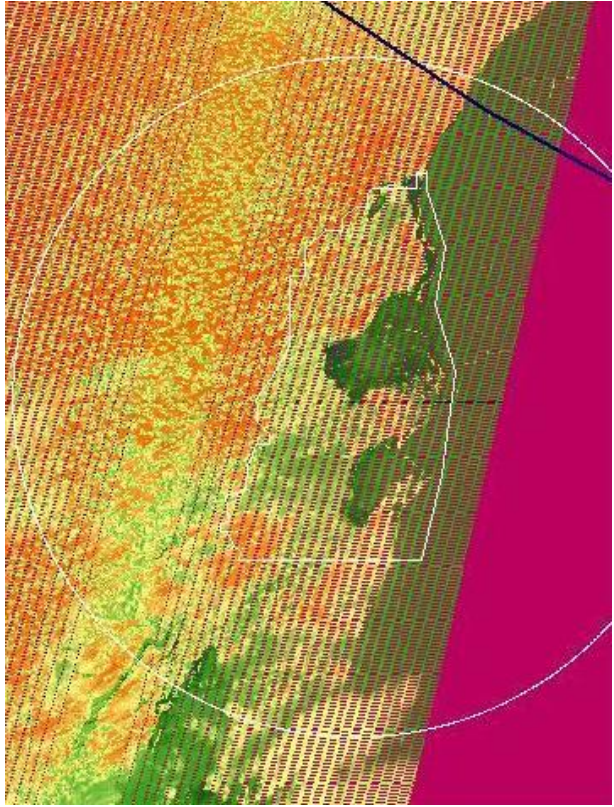

Fig. 8.  Landsat 7 NDVI composite of showing off nadir gap issue.  Sian Ka'an Boundary

and 100 km buffer are shown in white.  Hurricane Emily track shown in blue.

SPOT satellite imagery over the Sian Ka'an for Hurricanes Roxanne and Dolly is

nonexistent due to location and dates of occurrence.  Emily and Dean imagery was.  For

the Everglades study area, Katrina (2005) and Wilma (2005) were selected because of

their proximity to the study site and their dates of occurrence of coincided with the dates

Emily and Dean.  SPOT imagery for Katrina and Wilma was also available.

**4.2 SPOT Data Source**

 SPOT 20 m imagery was acquired from the Center for Southeast Tropical Advanced Remote Sensing (CSTARS) at the University of Miami (see http://cstars.rsmas.miami.edu for more information).  Six false color composites were used for the study, 2 for Dean, 2 for Emily, and 2 for Katrina and Wilma (Table 1).  Hurricanes Katrina and Emily were combined because of their closeness in proximity and date of occurrence.  Katrina occurred on August 25 and Wilma 24 October.

| | Image Date | Days Before/After Hurricane | % Cloud Cover per Scene | Satellite |
|---|---|---|---|---|
| **Sian Ka'an** | | | | |
| *Dean* | | | | |
| Before Landfall | 6/18/07 | 65 Days Before | 28%, 13%, 11% | 3 SPOT-2 Scenes |
| After Landfall | 11/15/07 | 86 Days After | 10%, 6% | 2 SPOT-4 Scenes |
| *Emily* | | | | |
| Before Landfall | 5/25/05 | 56 Days Before | 13%, 21% | 2 SPOT-4 Scenes |
| After Landfall | 8/1/05 | 14 Days After | 42%, 27% | 2 SPOT-4 Scenes |
| **Everglades** | | | | |
| *Katrina & Wilma* | | | | |
| Before Landfall | 3/2/05 | 176 Days Before Katrina | 0% | SPOT-4 Scenes |
| | | 236 Days Before Katrina | | |
| After Landfall | 10/27/05 | 63 Days After Katrina | 0% | SPOT-4 Scenes |
| | | 3 Days After Wilma | | |

Table 1.  SPOT Image Characteristics.

**4.3 Sian Ka'an Biosphere Reserve Imagery**

Fig. 9A shows a false color NDVI composite of Quintana Roo taken on 6/18/2007, 65 days before Dean made landfall.  The image consists of three SPOT 2 scenes that have mean cloud coverage of approximately 17.5%.  GIS software (Idrisi version 15, Andes Edition) was used to develop a raster mask to remove all clouded areas from analysis. Vector polygons were digitized based on land use categories-- cloud, water, urban, and vegetation, spectral signatures were developed for each band (1-3), and maximum likelihood supervised classification was conducted to classify the image assigning the value of "1" to clouded areas and "0" to all others, producing a functioning cloud mask raster.  This method was used for all composites that have cloud coverage.  Fig. 9B, on the right, is a false color NDVI composite of Quintana Roo taken on 11/15/2007, 86 days after Dean made landfall. The image consists of two SPOT 2 scenes that have a mean cloud coverage of approximately 8%.

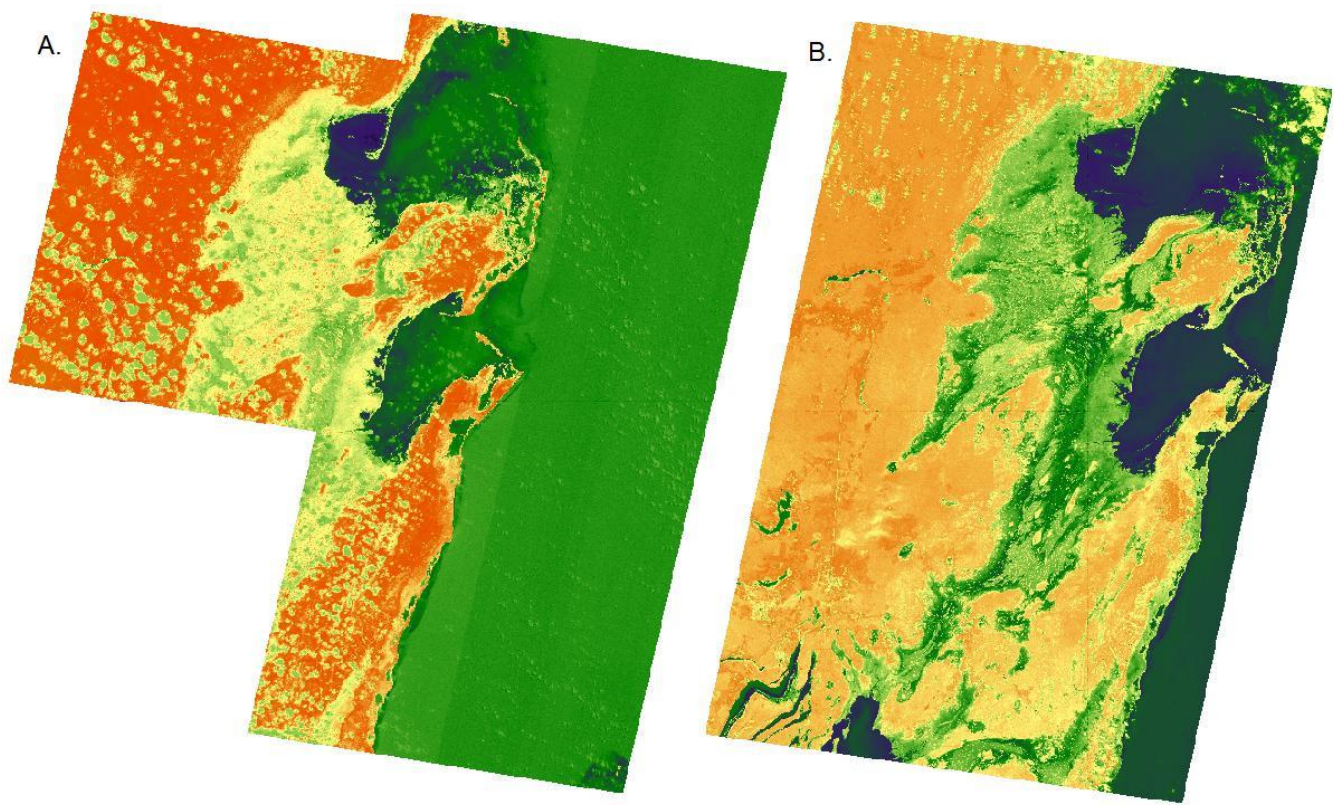

Figs. 9A and 9B. False color NDVI composites of the Quintana Roo area before and after Hurricane Dean made landfall, respectively.

Fig. 10A shows a false color NDVI composite of the northern area of the Sian Ka'an Biosphere Reserve taken on 5/25/2005, 56 days before Dean made landfall. The image consists of two SPOT 4 scenes that have mean cloud coverage of approximately 17%. Fig. 10B is a false color NDVI composite of the northern area of the Sian Ka'an Biosphere Reserve taken on 8/1/2005, 14 days after Dean made landfall. The image consists of two SPOT 4 scenes that have a mean cloud coverage of approximately 34.5%.

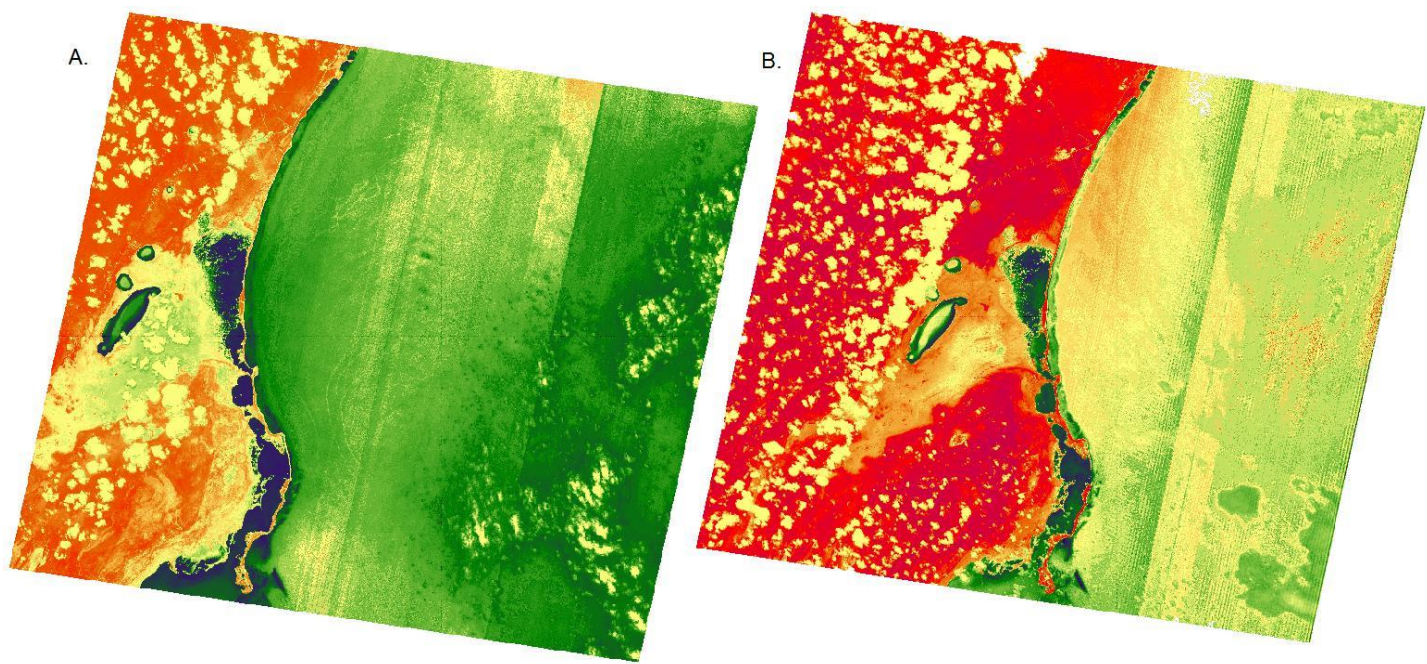

Figs. 10A and 10B. False color NDVI composites of the Quintana Roo area before and
after Hurricane Emily made landfall, respectively.

## 4.4 Everglades National Park Imagery

Figs. 11A and 11B show SPOT 4 imagery collected before the hurricanes in March 2005
and post hurricane season 27 October 2005.  Both figures 11A and 11B are cloud free.

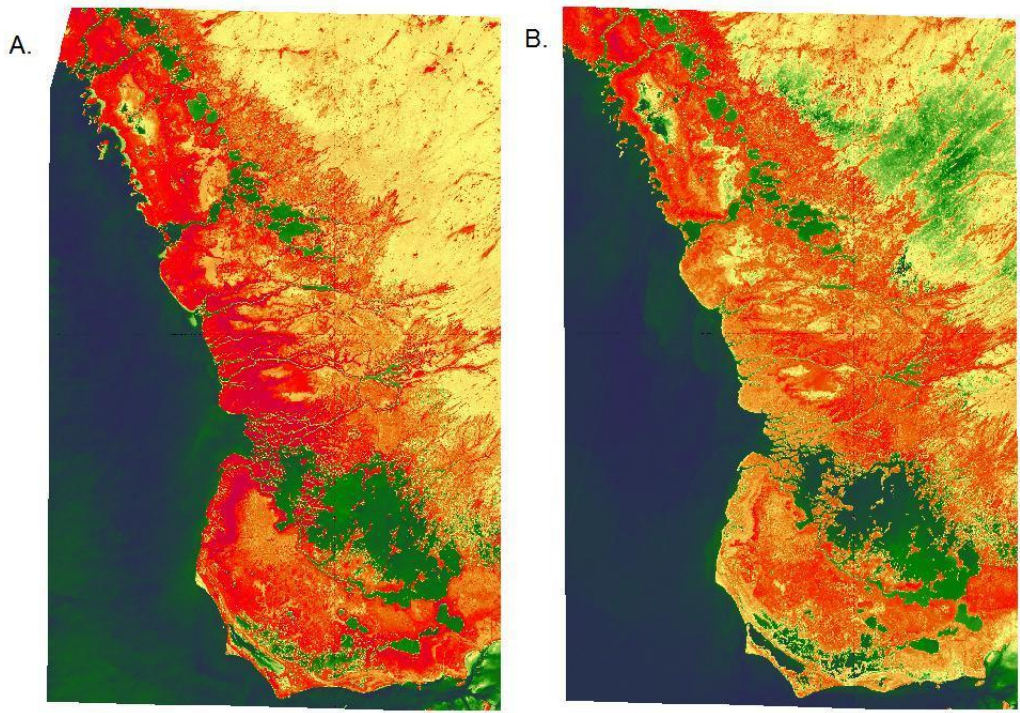

Figs. 11A and 11B. False color NDVI composites of the South Florida area before and after Hurricanes Wilma and Katrina made landfall, respectively.

## 4.5 Point Sampling

Mangrove areas were established base on a georeferenced vegetation map provided by the Amigos de Sian Ka'an.  Mangrove area polygons and sample points (Dean, n = 1567; Emily, n = 394; Katrina and Wilma, n = 644) were digitized for all sets of NDVI imagery using GIS software (Figs. 12-14).  Points were digitized manually out of consideration for the overlaid cloud mask raster, which is not incorporated in the random point generator processing.

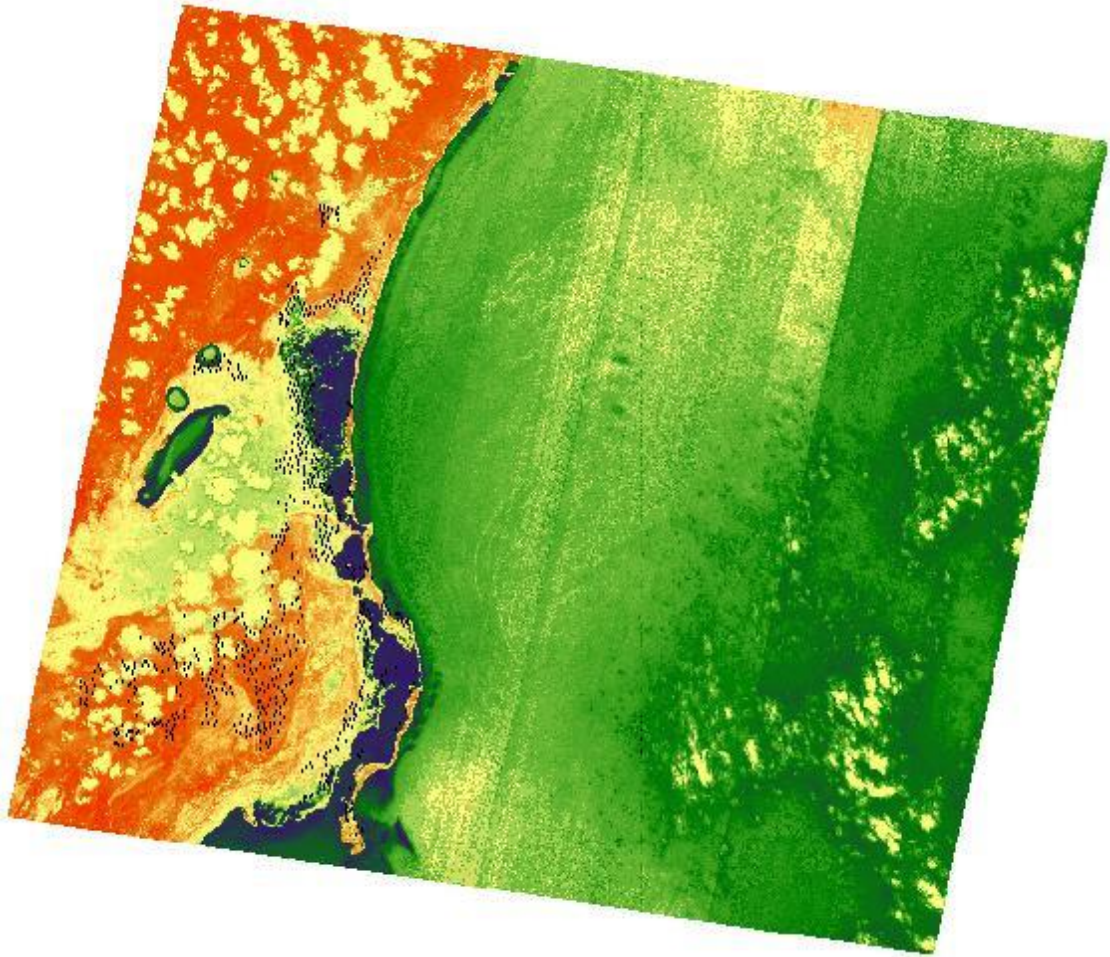

Fig. 12. Emily SPOT imagery with mangrove sample points shown as black dots.

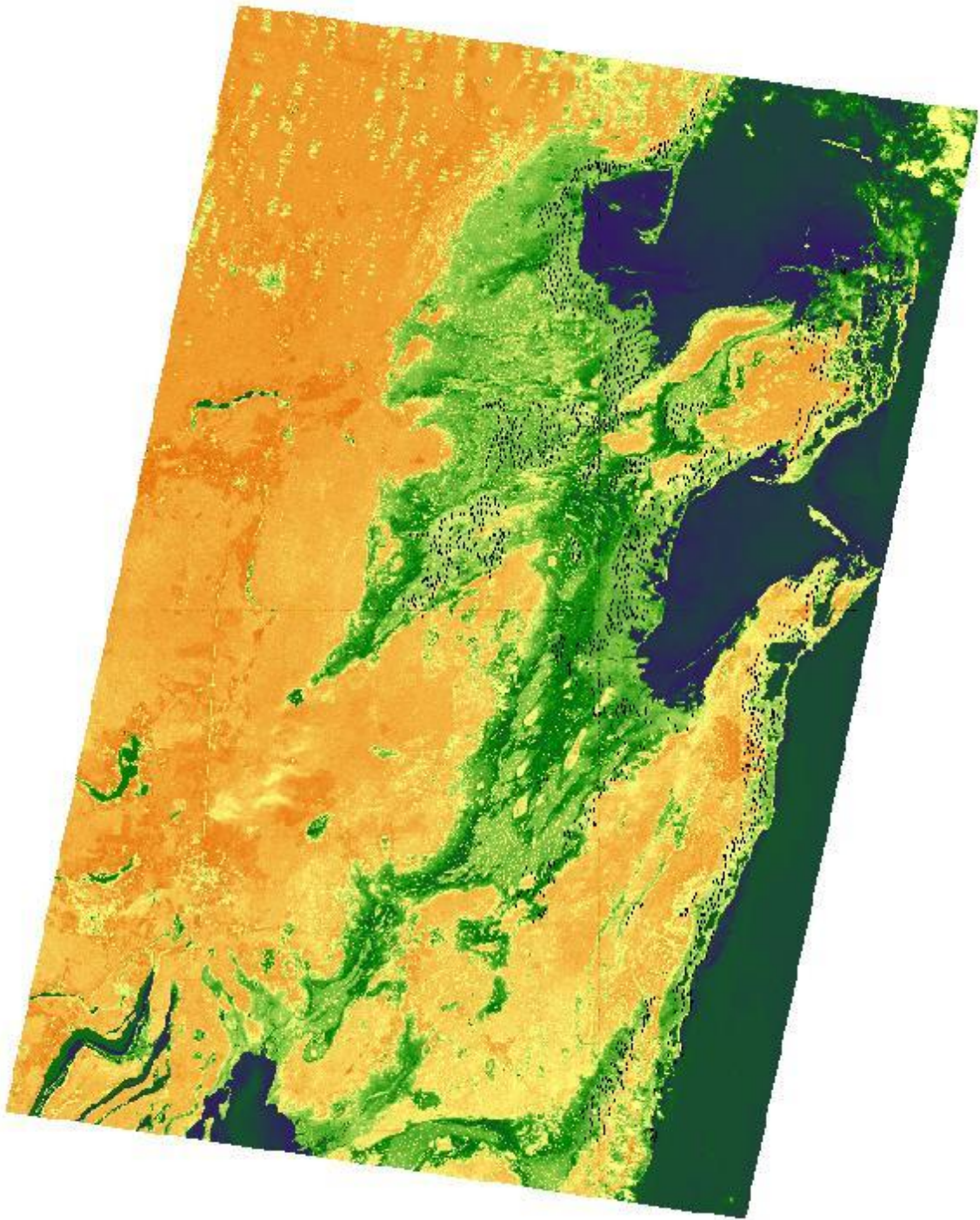

Fig. 13. Dean SPOT imagery with mangrove sample points shown as black dots.

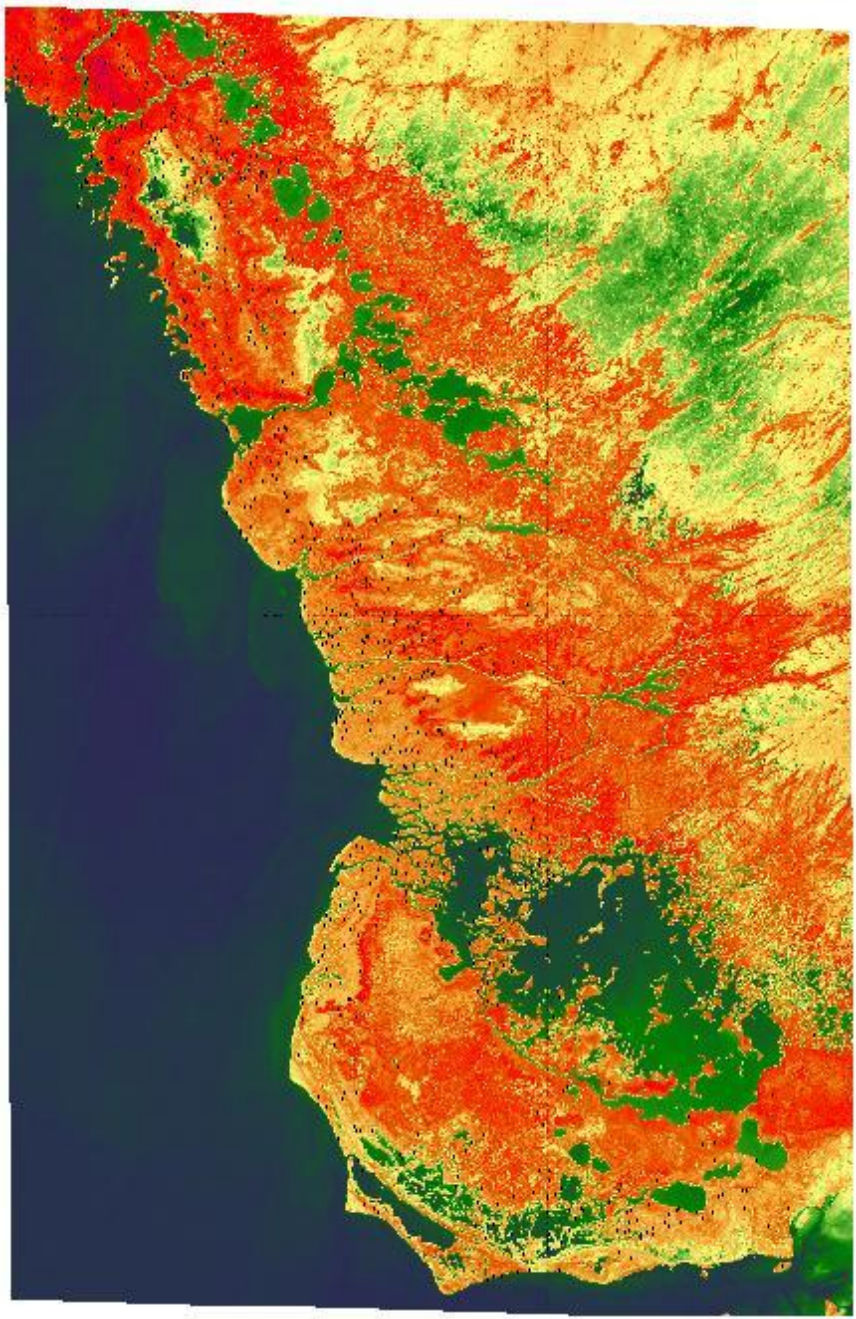

Fig. 14. Katrina and Wilma SPOT imagery with mangrove sample points shown as black dots.

Ripley's K Function was used to ensure that the manually sampled points were dispersed across the mangrove areas evenly.  This process was conducted using ArcGIS's Multi-Distance Spatial Cluster Analysis tool.  Other pattern analysis tools available in ArcGIS do not permit polygon constrained area analysis.  For example, the average nearest neighbor tool allows for the Cartesian extents, min/max x and min/max y, to be defined by a selected polygon's min/maxes, resulting in a quadrilateral area analysis.  Ripley's K Function allows for the spatial analysis to be constrained to the area of a selected polygon (Equation 3).  The specific area analysis has made Ripley's K useful for researchers to summarize vegetation patterns, test hypotheses about vegetation patterns, and fit vegetation models [33-35]. The Multi-Distance Spatial Cluster Analysis tool uses a common transformation of Ripley's K function shown below:

$$L(d) = \sqrt{\frac{A \sum_{i=1}^{N} \sum_{j=1, j \neq i}^{N} k(i,j)}{\pi N(N-1)}} \qquad (3)$$

where $A$ is area, $N$ is the number of points, $d$ is the distance and $k(i, j)$ is the weight, which (if there is no edge correction) is 1 when the distance between $i$ and $j$ is less than or equal to $d$ and 0 when the distance between $i$ and $j$ is greater than $d$. When edge correction is applied, the weight of $k(i,j)$ is modified slightly [33].

Figs. 15-17 shows the results of Ripley's K Function for each set of imagery graphically.  Each graph has two lines—expected and observed.  The expected line represents the random spatial pattern according constraints of the area (i.e. the mangrove area polygon).  The observed line is the actual observed spatial pattern.  The upper left area of the graph indicates statistically significant clustering at smaller distances and the lower right area indicates statistically significant dispersion at larger distances.  For each of the figures an expected to dispersed pattern is observed over different point distances.

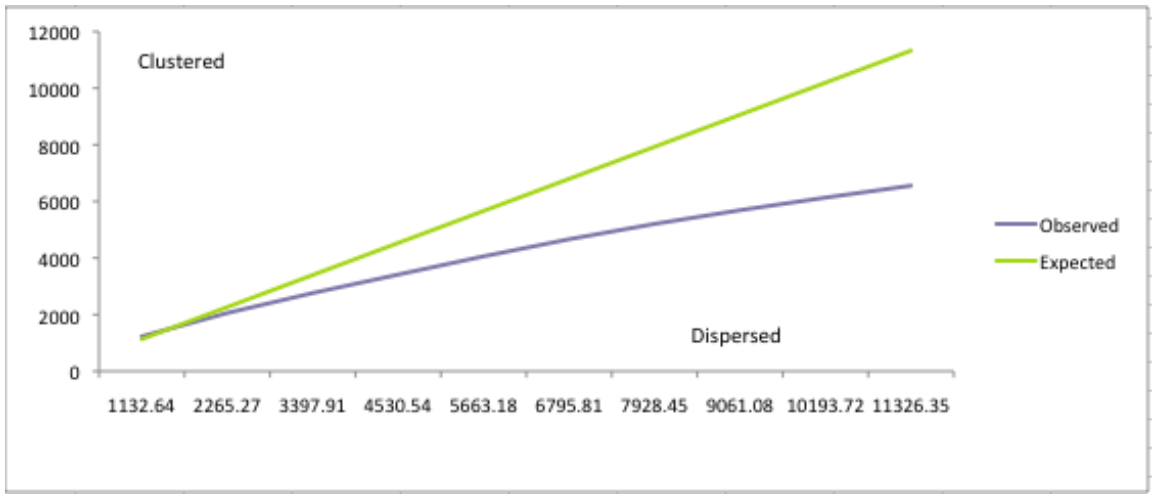

Fig. 15.  Dean mangrove point distribution over point distances.  X-axis units are point

distance meters and y-axis units are L(d), where d is distance in meters.

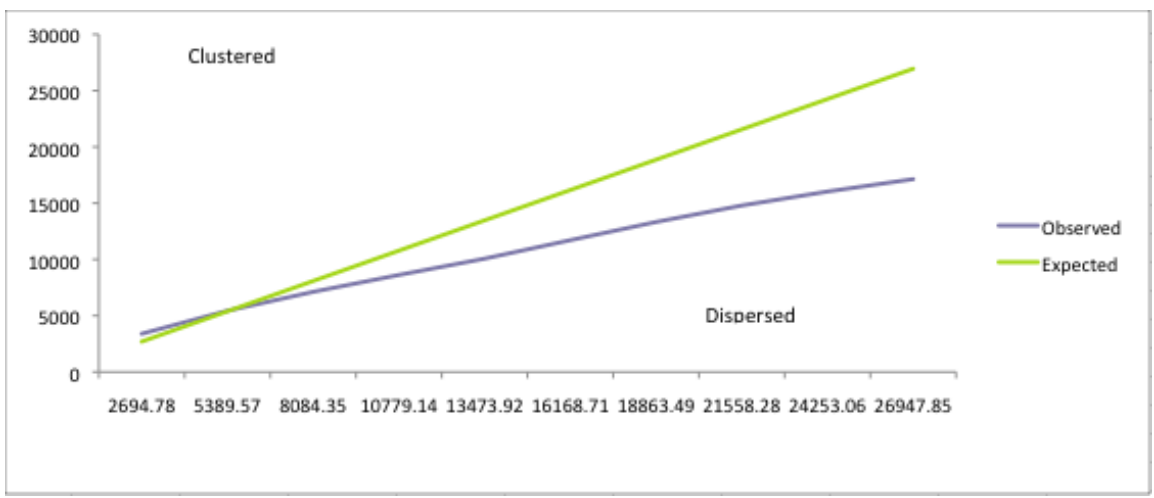

Fig. 16.  Emily mangrove point distribution over point distances. X-axis units are point

distance meters and y-axis units are L(d), where d is distance in meters.

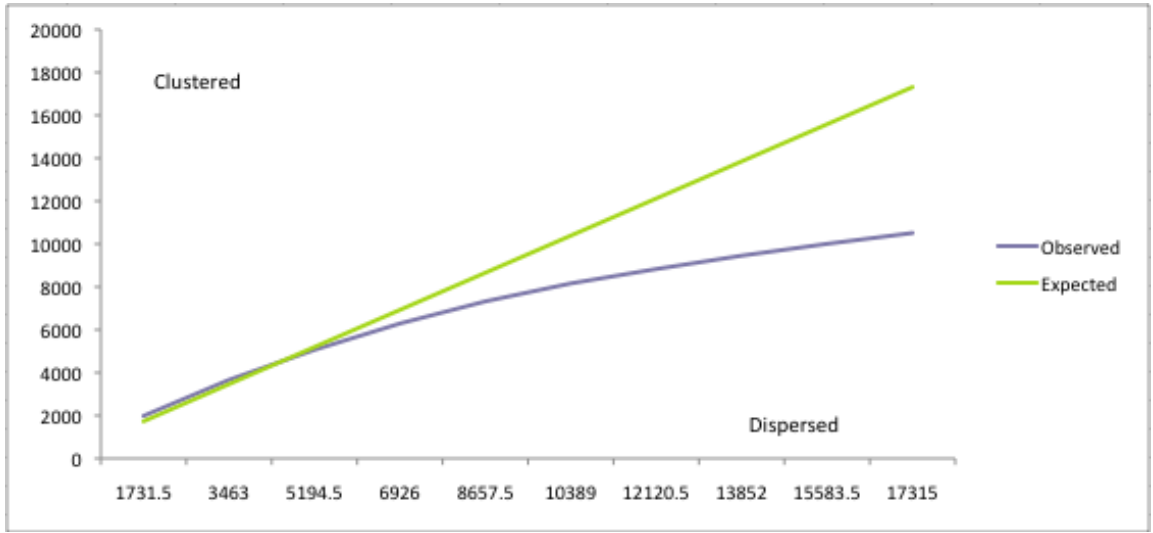

Fig. 17.  Katrina and Wilma mangrove point distribution over point distances. X-axis units are point distance meters and y-axis units are L(d), where d is distance in meters.

Each set of mangrove sample points were analyzed with the IDRISI GIS analysis extraction tool.  Two statistical processes were computed—arithmetic mean and standard deviation.

## 4.6 MODIS Data Source

Moderate Resolution Imaging Spectroradiometer (MODIS) sensors onboard the NASA Terra satellite obtains multispectral image data on a daily basis at various spatial and spectral resolutions. The availability of daily surface reflectance datasets enables the computation of spectral indices on a daily basis, which is well suited for studying land surface changes that occur in short periods of time [36].

MODIS vegetation indices imagery gridded to 500m resolution was obtained from http://modis.gsfc.nasa.gov/data/. To examine the effects of hurricanes as well as mangrove phenology, the NDVI and EVI time series for each site, the Sian Ka'an and the Everglades, was obtained from a series of 16-day MODIS composites (MOD 13 A2, Collection 4.0) from 2002-2008.

## 4.7 Univariate NDVI Differencing and Transect Sampling

Univariate NDVI differencing was employed to calculate SPOT imagery vegetation change for each set of images. This differencing method has shown to be effective way to detect vegetation change. Michener and Houhoulis [37] analyzed cross-referenced composite analysis, principle component analysis and univariate NDVI image differencing and found that the univariate NDVI differencing most accurately identified vegetation changes in their SPOT dataset. The image calculator software in IDRISI calculated NDVI differencing by subtracting each set's earlier image from the later image.

In order to understand the spatial trends in the univariate NDVI, 25 transects were digitized for each set of SPOT imagery (Figs. 18-20). All digitized transects start on the coast and end inland with consideration of the cloud masks.

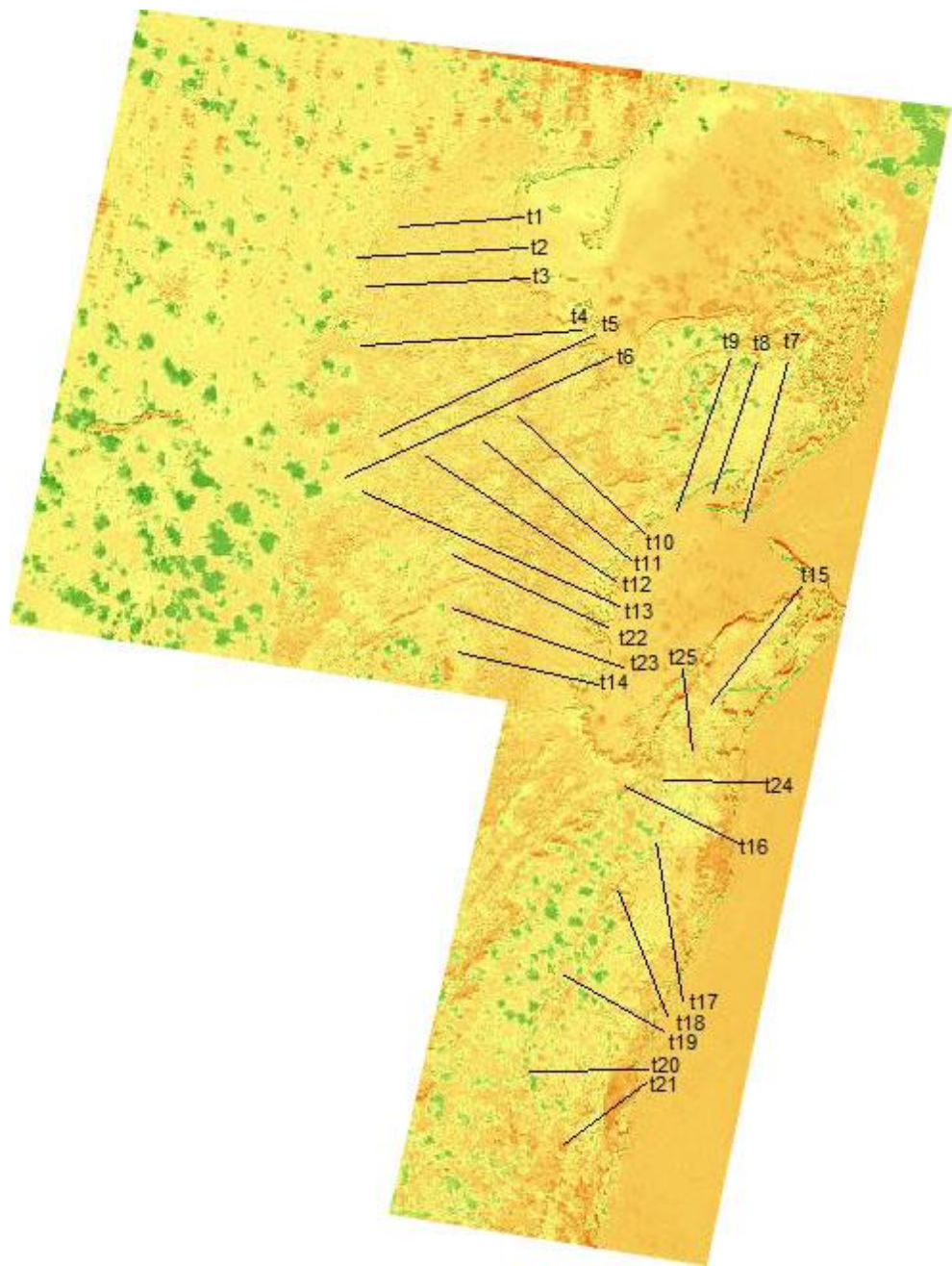

Fig. 18.  Univariate NDVI difference image (6/18/07 and 11/15/07 SPOT imagery for Dean) with 25 transects overlaid on a portion of the Sian Ka'an study area.

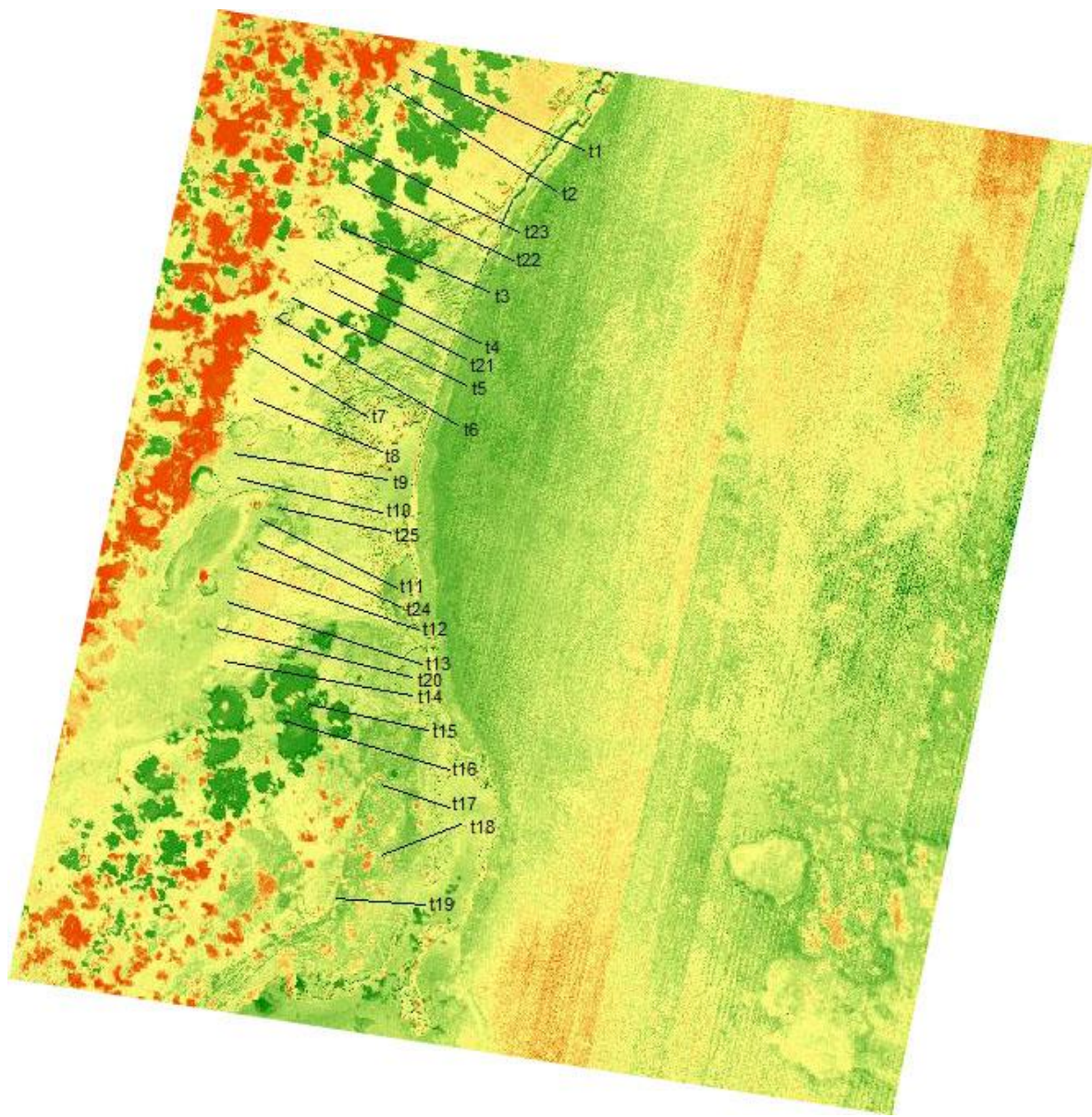

Fig. 19.   Univariate NDVI difference image (5/25/05 and 8/1/05 SPOT imagery for Emily) with 25 transects overlaid on a portion of the Sian Ka'an study area.

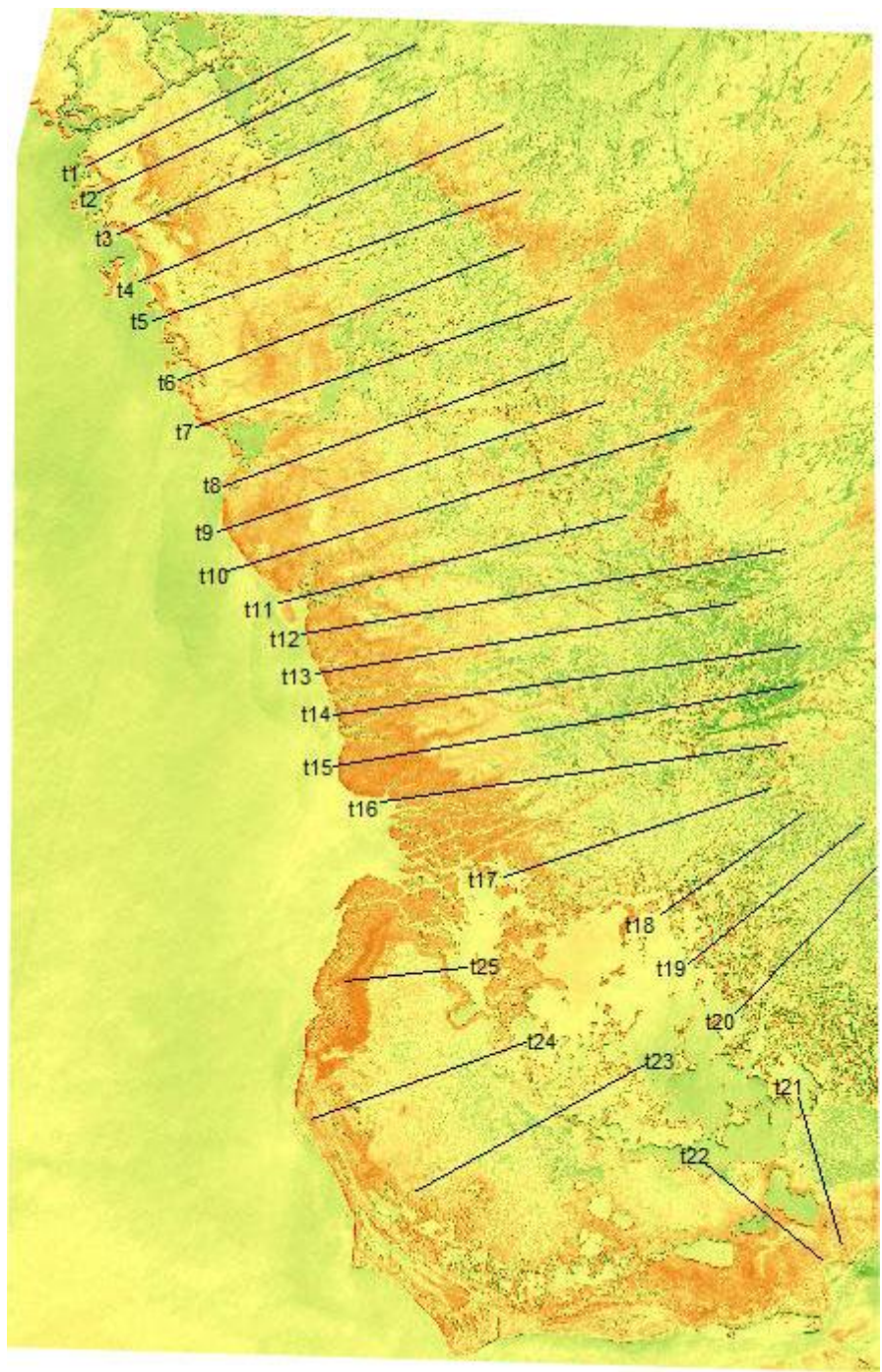

Fig. 20.  Univariate NDVI difference image (3/2/05 10/27/05 SPOT imagery for Katrina

and Wilma) with 25 transects overlaid on the Everglades study area.

# 5 RESULTS

## 5.1 Comparison of mean and standard deviation

The imagery from Hurricanes Dean, Katrina, and Wilma showed a significant drop in NDVI values after hurricane landfall.  The Emily results showed a significant increase in the mean NDVI value (Fig. 21).

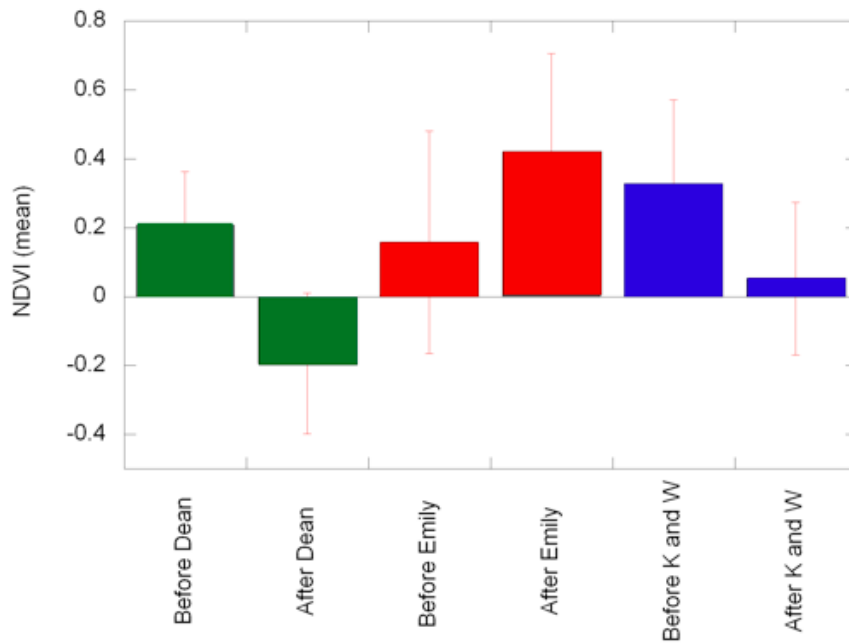

Fig. 21. Before and after mean NDVI values with vertical standard deviation lines overlaid.

All paired t-tests reported p-values less than .05 (p = .00 in all cases), which suggests that there is a statistically significant difference between the before and after NDVI values and that the differences between NDVI means are not likely due to change and are probably due to hurricane(s) landfall (Tables 3-5).

| Paired Differences | | | | | | | |
|---|---|---|---|---|---|---|---|
| | | | 95% Confidence Interval of the Difference | | | | |
| Mean | Std. Deviation | Std. Error Mean | Lower | Upper | t | df | Sig. (2-tailed) |
| .4358825 | .1838420 | .0206838 | .3947042 | .4770609 | 21.074 | 78 | .000 |

Table 3. Dean

| Paired Differences | | | | | | | |
|---|---|---|---|---|---|---|---|
| | | | 95% Confidence Interval of the Difference | | | | |
| Mean | Std. Deviation | Std. Error Mean | Lower | Upper | t | df | Sig. (2-tailed) |
| -0.26231 | 0.18163 | 0.00641 | -0.27490 | -0.24972 | -40.899 | 801 | 0.00 |

Table 4. Emily

| Paired Differences | | | | | | | |
|---|---|---|---|---|---|---|---|
| | | | 95% Confidence Interval of the Difference | | | | |
| Mean | Std. Deviation | Std. Error Mean | Lower | Upper | t | df | Sig. (2-tailed) |
| .2783760 | .2102066 | .0247731 | .2289799 | .3277721 | 11.237 | 71 | .000 |

Table 5. Katrina and Wilma

## 5.2 MODIS time series NDVI and EVI

MODIS NDVI and EVI data was gathered from a 6-year period spanning 2002 to 2008. Sequence plots show periodic seasonal variation occurring regularly in the Sian Ka'an and Everglades areas (Fig. 22).

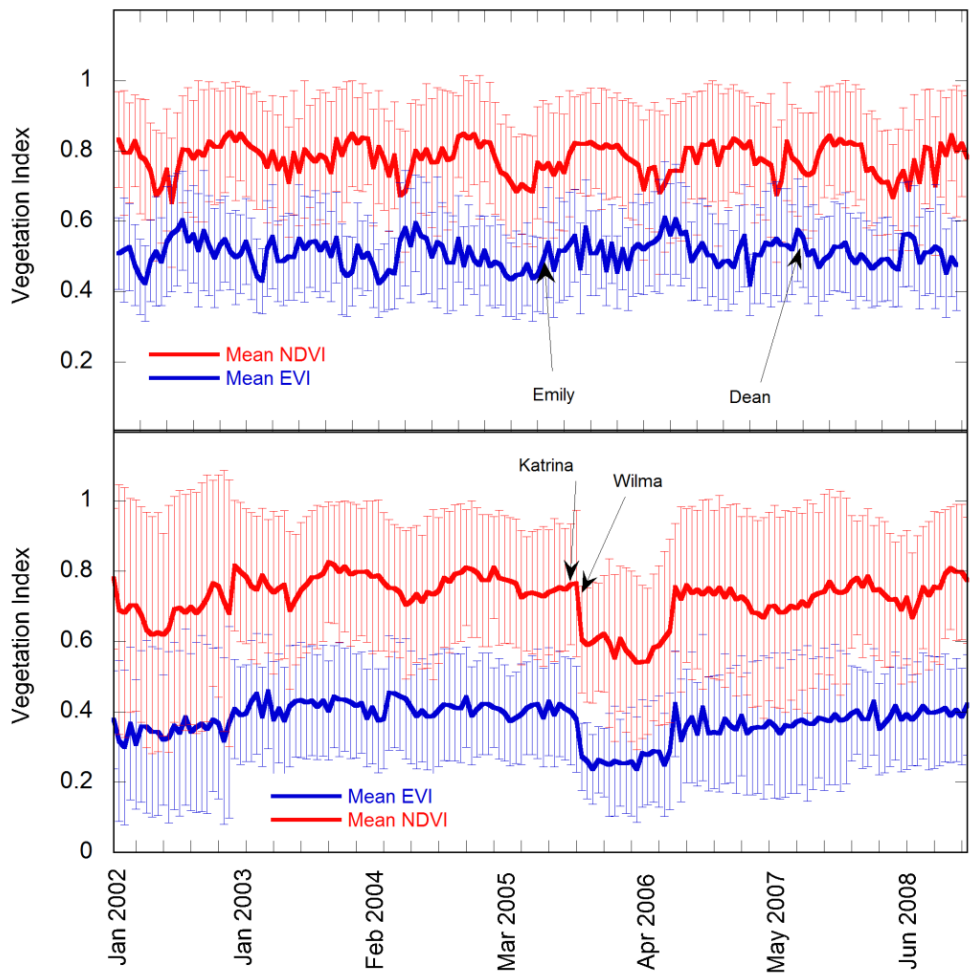

Fig. 22.  Upper: NDVI and EVI mean and standard deviation values of Sian Ka'an Biosphere Reserve over 6 year period.  Lower: NDVI and EVI mean and standard deviation values of Everglades National Park over same period.

The effects of Katrina and Wilma show a significant decline that departs from observed seasonal variation.  The mean NDVI and EVI values after their landfall are approximately -0.08 and -0.13, respectively.  Dean showed a less dramatic decrease in mean NDVI and EVI values of approximately -0.02 and -0.05, respectively.  The increase in mean NDVI and EVI for Emily is approximately 0.02 and 0.03, respectively.  The MODIS mean and standard deviation NDVI values and the SPOT mean and standard deviation NDVI values are in agreement.

## 5.3 NDVI differencing and transects

The transect difference NDVI values were graphed in groups of three based on proximity (Figs. 23-28).  The graphs illustrate the most NDVI volatility approximately between 0 – 4000 meters in all scenes.  This 4000 meter extent covers the shore area, which would take most of the storm surge effects.  Then, as we progress further inland (4000+ meters) NDVI values generally stabilize and show no further, discernable trend.

Hurricane Dean Transect Results

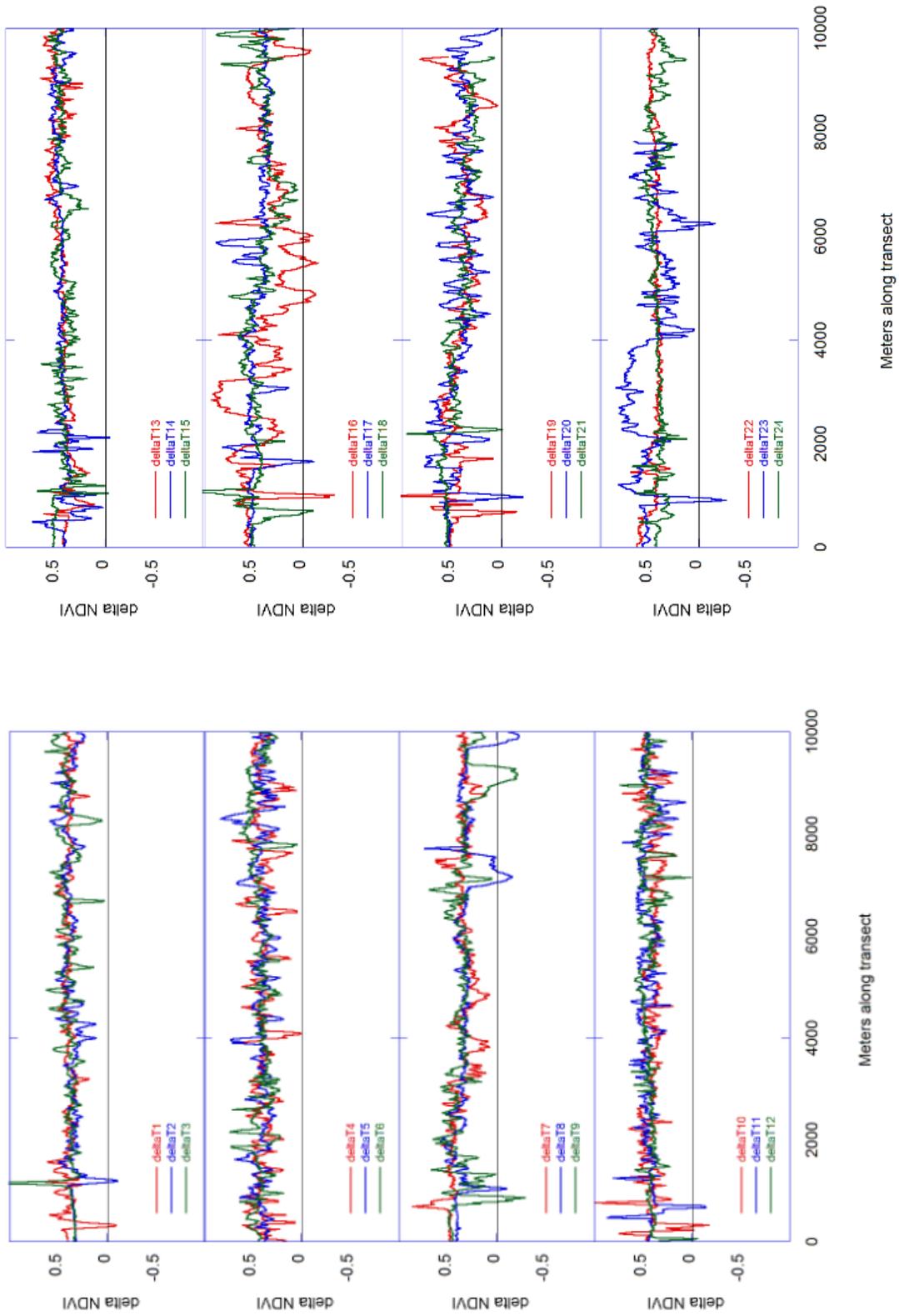

Fig. 23.

Fig. 24.

Hurricane Emily Transect Results

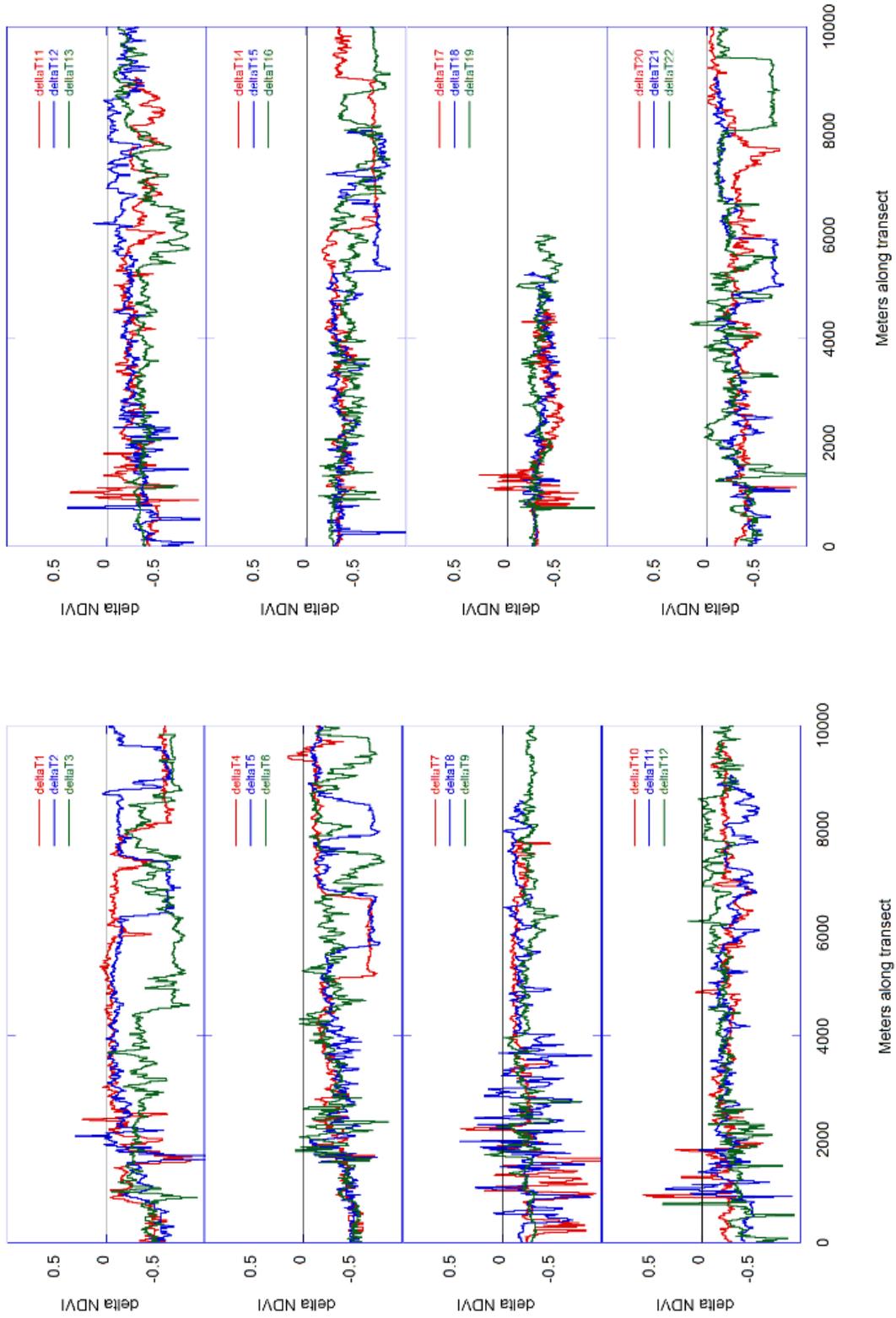

Fig. 25.

Fig. 26.

Hurricane Katrina and Wilma Transect Results

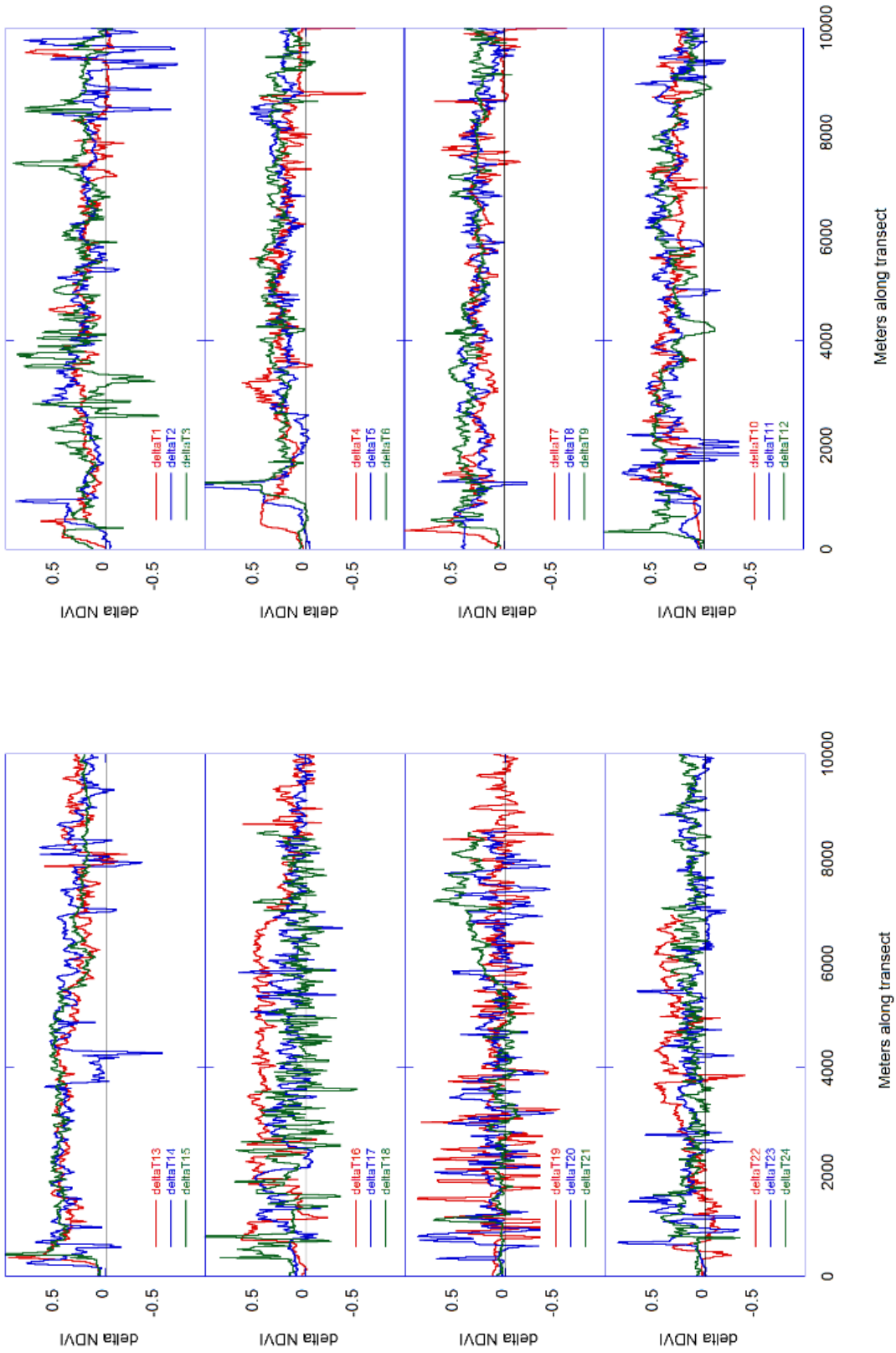

Fig. 27.

Fig. 28.

# 6 DISCUSSION AND CONCLUSIONS

Mangrove communities dominate the coastal regions of the Sian Ka'an Biosphere Reserve and the Everglades National Park. The mangrove species found within each protected area not only help form these communities but also help create habitats for a diverse and characteristic environment, which include numerous mangrove-dependent organisms including many fish species, invertebrates and vascular plants [38].

Despite the ecological significance of mangroves, mangrove swamps are an imperiled ecosystem. In South Florida and the Yucatan they are threatened by anthropogenic activities and natural disasters, namely hurricanes. To help curb the impact of anthropogenic activities the federal governments of Mexico and the United States have enacted preservation initiatives that designate sections of the mangrove communities as protected areas. But, even with legal protection, the mangrove communities experience damaging changes due to hurricanes and are further threatened by sea-level rise [39]. Our study has focused on the effects of two landfall hurricanes in the Sian Ka'an and two in the Everglades to better understand the impacts hurricanes have on the coastal mangrove communities.

Hurricanes can influence the structure of mangrove swamp communities through a variety of ways—wind damage, storm surges, and sediment deposition to name a few [40]. Immediate effects on mangrove species include alterations in canopy coverage, stem size, and vegetation density. The results of our study found that short-term changes in mangrove communities due to hurricanes can result in major damage, as seen with Hurricanes Wilma and Katrina, moderate damage, as seen with Hurricane Dean, and even growth, as seen with Hurricane Emily.

The literature focusing on hurricane impacts on mangrove communities is relatively well developed, but has consistently shown that hurricanes have detrimental effects on mangrove communities [41]. For example, Imbert *et al.* [42] found major damage occurring in the mangrove forests in Guadeloupe after Hurricane Hugo. Baldwin *et al.* [43] measured the change in plant community density in mangrove communities in South Florida after Hurricane Andrew. The results showed that severe damage and loss of canopy coverage had occurred and that regeneration of mangrove communites following hurricanes depend in large part on the density of seedlings that survive. Further, Ward *et al.* [40] explored the regenerative dynamics of the destroyed mangrove communities in southwest Florida after Hurricane Andrew. However, several studies [8, 44, 45] have demonstrated relatively rapid recovery of mangrove canopies even after extreme storm events.

The results from this study are generally consistent with those in the literature and and showed decreased canopy coverage after hurricane landfall except in the case of Hurricane Emily. The Sian Ka'an mangrove swamp communities experienced a rejuvenation of mangrove canopy coverage after Emily made landfall in the Yucatan.

According to the MODIS NDVI and EVI time series (Fig. 22), when Hurricane Emily made landfall NDVI and EVI values were at depressed levels for over a month, suggesting that mangrove canopies had been distressed from a dry season. The growth in mangrove canopies in the Sian Ka'an after Hurricane Emily could then be the result of added sediments needed by the mangrove canopies to achieve regentation [26].

Another consideration with regards to the atypical results shown by Hurricane Emily is the particular circumstances of each hurricane during landfall.  According the NOAA, in the Northern Hemisphere the highest winds of a hurricane will typically be on the right side.  In other words, if the hurricane is moving toward the west as Emily was, the strongest winds will be on the northern side, which is furthest from the Sian Ka'an Biosphere Reserve (Figure 6).  Therefore, although Emily reached category 5 status, once over the Yucatan its southern side may not have hit the Sian Ka'an with the same amount of force as the northern side in the Gulf of Mexico.

Each of the three satellite remote sensing and GIS approaches used in this study complemented each other.  The SPOT NDVI sample point calculations and the SPOT NDVI differencing both showed consistent results for each hurricane event.  In every event the MODIS NDVI and EVI time series results was consistent with the SPOT data. This was especially significant to confirm the SPOT NDVI results of Hurricane Emily, which resulted in an atypical increase in mangrove canopy coverage.  This study suggests that the sampling approaches allowed consistent monitoring of vegetative changes associated with mangrove areas following hurricanes.  Methodological refinements could include mangrove seed density via in situ measurements as suggested by Balwin *et al.* (2001) who found this as an important part of mangrove regeneration after hurricane landfall.

# References

[1] Odum, W.E. and C.C., McIvor. "Mangroves." In R.L. Myers, J.J. Ewel (Eds.), Ecosystems of Florida (pp. 517-548). University of Central Florida Press, Orlando (1990).

[2] Hogarth, P.J., The Biology of Mangroves. Oxford Oxford University Press, and New York, 228 pp. (1999).

[3] Saenger, P. and S.C. Snedaker, "Pantropical trends in mangrove above-ground biomass and annual litterfall," Oecologia 96, 293-299 (1993).

[4] Krauss, K.W., T.W. Doyle, R.R. Twilley, T.J. Smith III, K.R.T. Whelan, and J.T. Sullivan, "Woody debris in the mangrove forests of South Florida," Biotropica 37, 9-15 (2005).

[5] Piou, C., I.C. Feller, U. Berger, and F. Chi, "Zonation patterns of Belizean offshore mangrove forests 41 years after a catastrophic hurricane," Biotropica 38, 365-374 (2006).

[6] Muller, R.A. and G.W. Stone,"A climatology of tropical storm and hurricane strikes to enhance vulnerability prediction for the Southeast US coast," J. Coastal Res. 17, 949-956 (2001).

[7] Virmani, J.I. and R.H. Weisberg, "The 2005 hurricane season: An echo of the past or a harbinger of the future?" Geophys. Res. Lett. 33, L05707, doi:10.1029/2005GL025517 (2006).

[8] Baldwin, A., M. Egnotovich, M. Ford, and W. Platt, "Regeneration in fringe mangrove forests damaged by Hurricane Andrew," Plant Ecol. 157, 149-162 (2001).

[9] Blasco, F.T., T. Gauquelin, M. Rasolofoharinoro, J. Denis, M. Aizpuru, and V. Caldairou, "Recent advances in mangrove studies using remote sensing data," Marine and Freshwater Res. 49, 287-296 (1998).

[10] Diaz, B.M. and G.A. Blackburn, "Remote sensing of mangrove biophysical properties: evidence from a laboratory simulation of the possible effects of background variation on spectral vegetation indices." Int. J. Remote Sens. 24, 53-73 (2003).

[11] Murray, M.R., S.A. Zisman, P.A. Furley, D.M. Munro, J. Gibson, J. Ratter, S. Bridgewater, C.D. Minty, and C.J. Place, "The mangroves of Belize Part 1. Distribution, composition and classification," Forest Ecol. Manag. 174, 265-279 (2003).

[12] Vijay, V., R.S. Biradar, A.B. Inamdar, G. Deshmukhe, S. Baji, and M. Pikle, "Mangrove mapping and change detection around Mumbai (Bombay) using remotely sensed data," Ind. J. Marine Sci. 34, 310-315 (2005).

[13] Thampanya, U., J.E. Vermaat, S. Sinsakul, and N. Panapitukkul, "Coastal erosion and mangrove progradation of Southern Thailand," Estuarine Coast. Shelf Sci. 68, 75-85 (2006).

[14] Held, A., C. Ticehurst, L. Lymburner, and M. Williams, "High resolution mapping of tropical mangrove ecosystems using hyperspectral and radar remote sensing," Int. J. Remote Sens. 24, 2739-2759 (2003).

[15] Hirano, A., M. Madden and R.Welch, "Hyperspectral image data for mapping wetland vegetation," Wetlands 23, 436-448 (2003).

[16] Rodriguez, W., D.W. Urish, I.C. Feller, and R.M. Wright, "Relationships between frequency of ground exposure and forest cover in a mangrove island ecosystem," ATOLL Research Bulletin. 568, (2009).

[17] Muttitanon, W., Tripathi, N.K. (2005). Land use/land cover changes in the coastal zone of Ban Don Bay, Thailand using Landsat 5 TM data. International Journal of Remote Sensing, 26 (11), pp. 2311-2323.

[18] Sellers, P.J. (1985). Canopy reflectance, photosynthesis and transpiration. International Journal of Remote Sensing, 6 (8), pp. 1335-1372.

[19] Sellers, P.J., Berry, J.A., Collatz, G.J., Field, C.B., Hall, F.G. (1992). Canopy reflectance, photosynthesis, and transpiration. III. A reanalysis using improved leaf models and a new canopy integration scheme. Remote Sensing of Environment, 42 (3), pp. 187-216.

[20] Matsushita, B., Yang, W., Chen, J., Onda, Y., Qiu, G. Sensitivity of the Enhanced Vegetation Index (EVI) and Normalized Difference Vegetation Index (NDVI) to topographic effects: A case study in high-density cypress forest (2007) Sensors, 7 (11), pp. 2636-2651.

[21] Olmsted, I., Lopez, A. & Duran, R. (1990). La vegetacion de Sian Ka'an. In Navarro, D.& Robinson, J. (1990) op.cit. Pp 47-94.

[22] United Nations Environment Programme, World Conservation Monitoring Centre, World Heritage Site: Sian Ka'an. June 1987. Updated 5-1990,, 8-1995, 2-1996, April 2008.

[23] Alongi, D.M. The Energetics of Mangrove Forests. Springer Science. LaVergne, TN, USA. 2009.

[24] Mazzotti, F.J., Fling, H.E., Merediz, G., Lazcano, M., Lasch, C., Barnes, T. Conceptual ecological model of the Sian Ka'an Biosphere Reserve, Quintana Roo, Mexico (2005) Wetlands, 25 (4), pp. 980-997.

[25] Langdon D. Clough. Everglades National Park, United States. Encyclopedia of Earth. Eds. Cutler J. Cleveland. Washington, D.C. Environmental Information Coalition, National Council for Science and the Environment. United Nations Environment Programme-World Conservation Monitoring Centre. First published in the Encyclopedia of Earth January 16, 2009; Last revised August 18, 2009; Retrieved April 3, 2010.

[26] United Nations Environment Programme, World Conservation Monitoring Centre, World Heritage Site: Sian Ka'an. June 1987. Updated 1982. Updated 8-1986, 6-1987.10-1990, 6-1995, 5-1997,10-1998, 7-2002, 3 -2003, November. 2007.

[27] Pasch, R.J., E.S. Blake, H.D. Cobb III, and D.P. Roberts, "Tropical Cyclone Report: Hurricane Wilma, National Hurricane Center, Miami, FL, USA," (available on line at http://www.nhc.noaa.gov/2005atlan.shtml) (2006).

[28] Franklin, J. L., Tropical Cyclone Report: Hurricane Dean. National Hurricane Center. 31 January 2008.

[29] Franklin, J. L. and Daniel P. Brown., Tropical Cyclone Report: Hurricane Emily. National Hurricane Center. 10 March 2006.

[30] Richard D. Knabb, Jamie R. Rhome, and Daniel P. Brown. Tropical Cyclone Report: Hurricane Katrina.  20 December 2005.

[31] Pasch, R. J., Eric S. Blake, Hugh D. Cobb III, and David P. Roberts.   Tropical Cyclone Report Hurricane Wilma.  12 January 2006.

[32] National Oceanic and Atmospheric Administration.  Hurricane Basics, May 1999.

[33] Stoyan, D. and Penttinen, A. 2000, Recent applications of point process methods in forestry statistics. Statistical Science 15, 61-78.

[34] Peterson, C. J. and Squiers, E. R. 1995 An unexpected change in spatial pattern across 10 years in an Aspen-White Pine forest, Journal of Ecology 83, 847-855.

[35] Duncan, R. P. 1993. Testing for life historical changes in spatial patterns of four tropical tree species in

Westland, New Zealand, Journal of Ecology 81, 403-416.

[36] ESRI. Multi-Distance Spatial Cluster Analysis (Ripley's k-function) (Spatial Statistics). January 9, 2009.

[37] Pu, R., Gong, P., Tian, Y., Miao, X., Carruthers, R.I., Anderson, G.L. Using classification and NDVI differencing methods for monitoring sparse vegetation coverage: A case study of saltcedar in Nevada, USA. (2008) International Journal of Remote Sensing, 29 (14), pp. 3987-4011.

[38] Feller, I.C., Marsha Sitnik. Mangrove Ecology: A Manual for a Field Course. Smithsonian Institution. Washington, D.C., 1996.

[39] Ross, M.S., O'Brien, J.J., Ford, R.G., Zhang, K., Morkill, A. Disturbance and the rising tide: The challenge of biodiversity management on low-island ecosystems. (2009) Frontiers in Ecology and the Environment, 7 (9), pp. 471-478.

[40] Smith III, T.J., Anderson, G.H., Balentine, K., Tiling, G., Ward, G.A., Whelan, K.R.T. Cumulative impacts of hurricanes on Florida mangrove ecosystems: Sediment deposition, storm surges and vegetation (2009) Wetlands, 29 (1), pp. 24-34.

[41] Ward, G. A., T. J. Smith III, K. R. T. Whelan, and T. W. Doyle. 2006. Regional processes in mangrove ecosystems: spatial scaling relationships, biomass, and turnover rates following catastrophic disturbance. Hydrobiologia 569:517–27.

[42] Imbert, D., P. Labbe, and A. Rousteau. 1996. Hurricane damage and forest structure in Guadeloupe. Journal of Tropical Ecology 12:663–80.

[43] Baldwin, A., M. Egnotich, M. Ford, and W. Platt. 2001. Regeneration in fringe mangrove forests damaged by Hurricane Andrew. Plant Ecology 157:149–62.

[44] Ross, M.S., Ruiz, P.L., Sah, J.P., Reed, D.L., Walters, J., Meeder, J.F. Early post-hurricane stand development in Fringe mangrove forests of contrasting productivity. (2006) Plant Ecology, 185 (2), pp. 283-297.

[45] Sherman, R.E., Fahey, T.J., Martinez, P. Hurricane impacts on a mangrove forest in the Dominican Republic: Damage patterns and early recovery. (2001) Biotropica, 33 (3), pp. 393-408.